\documentclass[runningheads]{svmult}
\usepackage{makeidx}   
\usepackage{graphicx}  
\usepackage{subeqnar}  
\usepackage{multicol}  
\usepackage{cropmark}  
\usepackage{physprbb}  
\def\plb#1#2#3{{Phys. Lett. B}
{\textbf {#1}},~#2~(#3)}
\def\apjl#1#2#3{{Astrophys. J. Lett.}
{\textbf {#1}},~#2~(#3)}
\def\apj#1#2#3{{Astrophys. Journal}
{\textbf {#1}},~#2~(#3)}
\def\jetpl#1#2#3{{JETP Lett.}
{\textbf {#1}},~#2~(#3)}
\def\jetpsp#1#2#3{{JETP (Sov. Phys.)}
{\textbf {#1}},~#2~(#3)}
\def\spss#1#2#3{{Sov. Phys. -Solid State}
{\textbf {#1}},~#2~(#3)}
\def\jpa#1#2#3{{J. Phys. A}
{\textbf {#1}},~#2~(#3)}
\def\pr#1#2#3{{Phys. Reports}
{\textbf {#1}},~#2~(#3)}
\def\mnras#1#2#3{{Mon. Not. Roy. Astr. Soc.}
{\textbf {#1}},~#2~(#3)}
\def\n#1#2#3{{Nature}
{\textbf {#1}},~#2~(#3)}
\def\cmp#1#2#3{{Commun. Math. Phys.}
{\textbf {#1}},~#2~(#3)}
\def\prsla#1#2#3{{Proc. Roy. Soc. London A}
{\textbf {#1}},~#2~(#3)}
\def\ptp#1#2#3{{Prog. Theor. Phys.}
{\textbf {#1}},~#2~(#3)}
\def\prd#1#2#3{{Phys. Rev. D}
{\textbf {#1}},~#2~(#3)}
\def\prl#1#2#3{{Phys. Rev. Lett.}
{\textbf {#1}},~#2~(#3)}
\def\ibid#1#2#3{{ibid.}
{\textbf {#1}},~#2~(#3)}
\def\npb#1#2#3{{Nucl. Phys. B}
{\textbf {#1}},~#2~(#3)}
\def\jhep#1#2#3{{JHEP}
{\textbf {#1}},~#2~(#3)}
\def\anj#1#2#3{{Astron. J.}
{\textbf {#1}},~#2~(#3)}
\def\baas#1#2#3{{Bull. Am. Astron. Soc.}
{\textbf {#1}},~#2~(#3)}
\def\grg#1#2#3{{Gen. Rel. Grav.}
{\textbf {#1}},~#2~(#3)}
\def\stmp#1#2#3{{Springer Trac. Mod. Phys.}
{\textbf {#1}},~#2~(#3)}
\def\ijmpa#1#2#3{{Int. J. Mod. Phys. A}
{\textbf {#1}},~#2~(#3)}

\begin{document}
\title*{Inflationary Cosmology
}
\toctitle{Inflationary Cosmology
}
%
%
\titlerunning{Inflationary Cosmology}
%
\author{George Lazarides}
\authorrunning{George Lazarides}
%
%
\institute{Physics Division, School of Technology,
Aristotle University of Thessaloniki,
Thessaloniki 540 06, Greece}

\maketitle

\begin{abstract}
The big bang \index{big bang} model and the history of the
early universe according to the grand unified theories are
introduced. The shortcomings of big bang \index{big bang}
are discussed together with their resolution by inflationary
cosmology. Inflation, \index{inflation} the subsequent
oscillation and decay of the inflaton, \index{inflaton}
and the resulting `reheating' \index{reheating} of the
universe are studied. The density perturbations
\index{density perturbations} produced by inflation
\index{inflation} and the temperature fluctuations
\index{temperature fluctuations} of the cosmic background
radiation are discussed. The hybrid inflationary model is
described. Two `natural' extensions of this model which
avoid the disaster encountered in its standard realization
from the overproduction of monopoles \index{monopole}
are presented. Successful `reheating' \index{reheating}
satisfying the gravitino constraint
\index{gravitino constraint} takes place after the end of
inflation \index{inflation} in all three versions of
hybrid inflation. \index{hybrid inflation} Adequate
baryogenesis \index{baryogenesis} via a primordial
leptogenesis \index{leptogenesis} occurs consistently
with the solar and atmospheric neutrino oscillation
\index{neutrino oscillations} data. The primordial lepton
asymmetry \index{lepton asymmetry} is turned partly into
baryon asymmetry \index{baryon asymmetry} via the
sphalerons \index{sphaleron} which are summarized.
\end{abstract}

\section{Introduction}
\label{sec:introduction}

\par
The discovery of the cosmic microwave background radiation
(CMBR) \index{CMBR} in 1964 together with the observed
Hubble expansion of the universe had established hot big
bang \index{big bang} cosmology \cite{wkt} as a viable
model of the universe. The success of the theory of
nucleosynthesis \index{nucleosynthesis} in reproducing the
observed abundance pattern of light elements together with
the proof of the black body character of the CMBR
\index{CMBR} then imposed hot big bang \index{big bang}
as the standard cosmological model. This model combined with
grand unified theories (GUTs) \cite{ggps} \index{GUT} of
strong, weak and electromagnetic interactions provides an
appropriate framework for discussing the very early stages of
the universe evolution.

\par
Despite its great successes, the standard big bang
\index{big bang} (SBB) \index{SBB} cosmological model had
a number of long-standing shortcomings. One of them is the
so-called horizon \index{horizon} problem. The CMBR
\index{CMBR} which we receive now has been emitted from
regions of the sky which never communicated causally before
sending light to us. The question then arises how come the
temperature of the black body radiation from these regions is
so finely tuned as the measurements of the cosmic background
explorer (COBE) \index{COBE} \cite{cobe} show. Another
issue is the flatness problem. \index{flatness problem} The
present universe appears almost flat. This requires that, in
its early stages, the universe was flat with a great accuracy,
which needs some explanation. Also, combined with GUTs
\index{GUT} which predict the existence of superheavy
magnetic monopoles \index{monopole} \cite{monopole}, the
SBB \index{SBB} model leads \cite{preskill} to a
cosmological catastrophe due to the overproduction of these
monopoles. \index{monopole} Finally, the model does not
explain the origin of the small density perturbations
\index{density perturbations} required for the structure
formation \index{structure formation} in the universe
\cite{structure} and the generation of the observed
\cite{cobe} temperature fluctuations
\index{temperature fluctuations} in the CMBR. \index{CMBR}

\par
Inflation \cite{guth,lindebook} \index{inflation} offers an
elegant solution to all these problems of the SBB \index{SBB}
model. The idea behind inflation \index{inflation} is that,
in the early universe, a real scalar field (the inflaton)
\index{inflaton} was displaced from its vacuum value. If
the potential energy density of this field happens to be quite
flat, the roll-over of the field towards the vacuum can be
very slow for a period of time. During this period, the energy
density is dominated by the almost constant potential energy
density of the inflaton. \index{inflaton} As a consequence,
the universe undergoes a period of quasi-exponential expansion,
which can readily solve the horizon \index{horizon} and
flatness problems \index{flatness problem} by stretching the
distance over which causal contact is established and
reducing any pre-existing curvature in the universe. It can
also dilute adequately the GUT \index{GUT} magnetic monopoles.
\index{monopole} Moreover, it provides us with the primordial
density perturbations \index{density perturbations} which are
necessary for explaining the large scale structure formation
\index{structure formation} in the universe \cite{structure}
as well as the temperature fluctuations
\index{temperature fluctuations} observed in the CMBR.
\index{CMBR} Inflation \index{inflation} can be easily
incorporated in GUTs. \index{GUT} It occurs during the GUT
\index{GUT} phase transition at which the GUT \index{GUT}
gauge symmetry breaks by the vacuum expectation value (vev)
of a Higgs field, which also plays the role of the inflaton.
\index{inflaton}

\par
After the end of inflation, \index{inflation} the inflaton
\index{inflaton} enters into an oscillatory phase about the
vacuum. The oscillations are damped because of the dilution of
the field energy density caused by the expansion of the
universe and the decay of the inflaton \index{inflaton} into
`light' matter. The radiation energy density generated by
the inflaton \index{inflaton} decay eventually dominates over
the field energy density and the universe returns to a normal
big bang \index{big bang} type evolution. The cosmic
temperature at which this occurs is historically called `reheat'
temperature \index{`reheat' temperature} although there is
actually neither supercooling nor reheating \index{reheating}
of the universe \cite{reheat}.

\par
An important disadvantage of the early realizations of inflation
\index{inflation} is that they require tiny coupling constants
in order to reproduce the COBE \index{COBE} measurements on the
CMBR. \index{CMBR} To solve this `naturalness' problem, the
hybrid inflationary scenario has been introduced \cite{hybrid}.
The basic idea was to use two real scalar fields instead of one
that was normally used. One field may be a gauge non-singlet and
provides the `vacuum' energy density which drives inflation,
\index{inflation} while the other is the slowly varying field
during inflation. \index{inflation} This splitting of roles
between two fields allows us to reproduce the temperature
fluctuations \index{temperature fluctuations} of the CMBR
\index{CMBR} with `natural' (not too small) values of the
relevant parameters in contrast to previous realizations of
inflation. \index{inflation} Hybrid inflation,
\index{hybrid inflation} although initially introduced in
the context of non-supersymmetric GUTs, \index{GUT} can be
`naturally' incorporated \cite{lyth,dss} in supersymmetric
(SUSY) \index{SUSY} GUTs. \index{GUT}

\par
Unfortunately, the GUT \index{GUT} monopole \index{monopole}
problem reappears in hybrid inflation. \index{hybrid inflation}
The termination of inflation, \index{inflation} in this case,
is abrupt and is followed by a `waterfall' regime during which
the system falls towards the vacuum manifold and starts
performing damped oscillations about it. If the vacuum manifold
is homotopically non-trivial, topological defects will be
copiously formed \cite{smooth} by the Kibble mechanism
\index{Kibble mechanism} \cite{kibble} since the system can
end up at any point of this manifold with equal probability. So
a cosmological disaster is encountered in the hybrid inflationary
models which are based on a gauge symmetry breaking predicting
the existence of magnetic monopoles. \index{monopole}

\par
One idea \cite{smooth,jean,shi} for solving the monopole
\index{monopole} problem of SUSY \index{SUSY} hybrid inflation
\index{hybrid inflation} is to include into the standard
superpotential for hybrid inflation \index{hybrid inflation}
the leading non-renormalizable term. This term cannot be
excluded by any symmetries and, if its dimensionless
coefficient is of order unity, can be comparable with the
trilinear coupling of the standard superpotential (whose
coefficient is $\sim 10^{-3}$). Actually, we have two
options. We can either keep \cite{jean} both these terms or
remove \cite{smooth,shi} the trilinear term by imposing an
appropriate discrete symmetry and keep only the leading
non-renormalizable term. The pictures which emerge in the two
cases are quite different. However, they share an important
common feature. The GUT \index{GUT} gauge group is already
broken during inflation \index{inflation} and thus no
topological defects can form at the end of inflation.
\index{inflation} Consequently, the monopole
\index{monopole} problem is solved.

\par
A complete inflationary scenario should be followed by a
successful `reheating' \index{reheating} satisfying the
gravitino constraint \index{gravitino constraint}
\cite{gravitino} on the `reheat' temperature,
\index{`reheat' temperature}
$T_r\stackrel{_{<}}{_{\sim}}10^9~{\rm GeV}$, and
generating the observed baryon asymmetry
\index{baryon asymmetry} of the universe (BAU). \index{BAU}
In hybrid inflationary models, it is \cite{lepto} generally
preferable to generate the BAU \index{BAU} by first
producing a primordial lepton asymmetry \cite{leptogenesis}
\index{lepton asymmetry} which is then partly converted
into baryon asymmetry \index{baryon asymmetry} by the
non-perturbative electroweak sphaleron \index{sphaleron}
effects \cite{dimopoulos,sphaleron}. Actually, in many
specific models, this is the only way to generate the BAU
\index{BAU} since the inflaton \index{inflaton} decays
into right handed neutrino superfields. The subsequent decay
of these superfields into lepton (antilepton) and electroweak
Higgs superfields can only produce a lepton asymmetry.
\index{lepton asymmetry} Successful `reheating'
\index{reheating} can be achieved \cite{jean,shi} in
hybrid inflationary models in accord with the experimental
requirements from solar and atmospheric neutrino oscillations
\index{neutrino oscillations} and with `natural' values of
parameters.

\section{The Big Bang \index{big bang} Model}
\label{sec:bigbang}

\par
We will start with an introduction to the salient features of
the SBB \index{SBB} model \cite{wkt} and a summary of the
history of the early universe in accordance to GUTs.
\index{GUT}

\subsection{Hubble Expansion}
\label{subsec:hubble}

\par
For cosmic times $t\stackrel{_{>}}{_{\sim }} t_{P}\equiv
M_{P}^{-1}\sim 10^{-44}~{\rm{sec}}$ ($M_{P}=1.22\times
10^{19}~{\rm{GeV}}$ is the Planck scale) after the big bang,
\index{big bang} the quantum fluctuations of gravity cease
to exist. Gravitation can then be adequately described by
classical relativity. Strong, weak and electromagnetic
interactions, however, require relativistic quantum field
theoretic treatment and are described by gauge theories.

\par
An important principle, on which SBB \index{SBB} is based,
is that the universe is homogeneous and isotropic. The
strongest evidence for this {\it cosmological principle}
is the observed \cite{cobe} isotropy of the CMBR.
\index{CMBR} Under this assumption, the four dimensional
space-time is described by the Robertson-Walker metric
\begin{equation}
ds^{2}=-dt^{2}+
a^{2}(t)\left[\frac{dr^{2}}{1-kr^2}
+r^{2}(d\theta^{2}+\sin^{2}\theta~d\varphi^{2})
\right],
\label{eq:rw}
\end{equation}
where $r$, $\varphi$ and $\theta$ are `comoving' polar
coordinates, which remain fixed for objects that have no
other motion than the general expansion of the universe.
$k$ is the `scalar curvature' of the 3-space and $k=0$,
$>0$ or $<0$ corresponds to flat, closed or open universe.
The dimensionless parameter $a(t)$ is the `scale factor'
of the universe and describes cosmological expansion. We
normalize it by taking $a_{0}\equiv a(t_{0})=1$, where
$t_{0}$ is the present cosmic time.

\par
The `instantaneous' radial physical distance is given by
\begin{equation}
R=a(t)\int_{0}^{r}\frac{dr}{(1-kr^{2})^{1/2}}~\cdot
\label{eq:dist}
\end{equation}
For flat universe ($k=0$), $\bar{R}=a(t)\bar{r}$
($\bar{r}$ is a `comoving' and  $\bar{R}$ a physical
vector in 3-space) and the velocity of an object is
\begin{equation}
\bar{V}=\frac{d\bar{R}}{dt}=\frac{\dot{a}}{a}\bar{R}
+a\frac{d\bar{r}}{dt}~,
\label{eq:velocity}
\end{equation}
where overdots denote derivation with respect to cosmic
time. The second term in the right hand side (rhs) of
this equation is the so-called `peculiar velocity',
$\bar{v}=a(t)\dot{\bar{r}}$, of the object, i.e., its
velocity with respect to the `comoving' coordinate
system. For $\bar{v}=0$, (\ref{eq:velocity}) becomes
\begin{equation}
\bar{V}=\frac{\dot{a}}{a}\bar{R}\equiv H(t)\bar{R}~,
\label{eq:hubblelaw}
\end{equation}
where $H(t)\equiv\dot{a}(t)/a(t)$ is the Hubble
parameter. This is the well-known Hubble law asserting
that all objects run away from each other
with velocities proportional to their distances and is
the first success of SBB \index{SBB} cosmology.

\subsection{Friedmann Equation \index{Friedmann equation}}
\label{subsec:friedmann}

\par
Homogeneity and isotropy of the universe imply that the
energy momentum tensor takes the diagonal form
$(T_{\mu}^{~\nu})={\rm{diag}}(-\rho, p, p, p)$,
where $\rho$ is the energy density of the universe and
$p$ the pressure. Energy momentum conservation
(${T_{\mu~;\nu}^{~\nu}}=0$) then takes the form of the
continuity equation
\begin{equation}
\frac{d\rho}{dt}=-3H(t)(\rho+p)~,
\label{eq:continuity}
\end{equation}
where the first term in the rhs describes the dilution of
the energy due to the expansion of the universe and the
second term corresponds to the work done by pressure.
Equation (\ref{eq:continuity}) can be given
the following more transparent form
\begin{equation}
d\left(\frac{4\pi}{3}a^{3}\rho\right)=
-p~4\pi a^{2}da~,
\label{eq:cont}
\end{equation}
which indicates that the energy loss of a `comoving'
sphere of radius $\propto a(t)$ equals the work
done by pressure on its boundary as it expands.

\par
For a universe described by the metric in (\ref{eq:rw}),
Einstein's equations
\begin{equation}
R_{\mu}^{~\nu}-\frac{1}{2}~\delta_{\mu}^{~\nu}R=
8\pi G~T_{\mu}^{~\nu},
\label{eq:einstein}
\end{equation}
where $R_{\mu}^{~\nu}$ and $R$ are the Ricci tensor
and scalar curvature and $G\equiv M_{P}^{-2}$ is the
Newton's constant, lead to the Friedmann equation
\index{Friedmann equation}
\begin{equation}
H^{2}\equiv \left(\frac{\dot{a}(t)}{a(t)}
\right)^{2}=\frac{8\pi G}{3}\rho-\frac{k}{a^{2}}~\cdot
\label{eq:friedmann}
\end{equation}

\par
Averaging $p$, we can write
$\rho+p=(1+w)\rho\equiv\gamma\rho$ and
(\ref{eq:continuity}) becomes $\dot{\rho}=
-3H\gamma\rho$, which gives $d \rho/\rho=-3 \gamma da/a$
and $\rho \propto a^{-3 \gamma}$. For a universe
dominated by pressureless matter, $p=0$ and, thus,
$\gamma=1$, which gives $\rho\propto a^{-3}$. This is
interpreted as mere dilution of a fixed number of particles
in a `comoving' volume due to the cosmological expansion.
For a radiation dominated universe, $p=\rho/3$ and, thus,
$\gamma=4/3$, which gives $\rho\propto a^{-4}$. Here, we
get an extra factor of $a(t)$ due to the red-shifting of
all wave lengths by the expansion. Substituting
$\rho \propto a^{-3 \gamma}$ in (\ref{eq:friedmann})
with $k=0$, we get $\dot{a}/a \propto a^{-3 \gamma/2}$ and,
thus, $a(t)\propto t^{2/3\gamma}$. Taking into account that
$a(t_{0})=1$, this gives
\begin{equation}
a(t)=(t/t_{0})^{2/3\gamma}.
\label{eq:expan}
\end{equation}
For a matter dominated universe, we get the expansion law
$a(t)=(t/t_{0})^{2/3}$. `Radiation', however, expands as
$a(t)=(t/t_{0})^{1/2}$.

\par
The early universe is radiation dominated and its energy
density is
\begin{equation}
\rho=\frac{\pi^{2}}{30}\left(N_{b}+\frac{7}{8}N_{f}
\right)T^{4}\equiv c~T^{4},
\label{eq:boltzman}
\end{equation}
where $T$ is the cosmic temperature and $N_{b(f)}$
the number of massless bosonic (fermionic) degrees of
freedom. The quantity $g_{*}=N_{b}+(7/8)N_{f}$ is
called effective number of massless degrees of freedom.
The entropy density is
\begin{equation}
s= \frac{2\pi^{2}}{45}~g_{*}~T^{3}.
\label{eq:entropy}
\end{equation}
Assuming adiabatic universe evolution, i.e., constant
entropy in a `comoving' volume ($sa^{3}={\rm{constant}}$),
we obtain $aT={\rm{constant}}$. The
temperature-time relation during radiation dominance is then
derived from (\ref{eq:friedmann}) (with $k=0$):
\begin{equation}
T^{2}=\frac{M_{P}}{2(8\pi c/3)^{1/2}t}~\cdot
\label{eq:temptime}
\end{equation}
Classically, the expansion starts at $t=0$ with $T=\infty$
and $a=0$. This initial singularity is, however, not
physical since general relativity fails for
$t\stackrel{_{<}}{_{\sim }}t_{P}$ (the Planck time). The
only meaningful statement is that the universe, after a
yet unknown initial stage, emerges at $t\sim t_{P}$ with
$T\sim M_{P}$.

\subsection{Important Cosmological Parameters}
\label{subsec:parameter}

\par
The most important parameters describing the expanding
universe are:
\begin{list}
\setlength{\rightmargin=0cm}{\leftmargin=0cm}
\item[{\bf i.}] The present value of the Hubble parameter
(known as Hubble constant) $H_{0}\equiv
H(t_{0})=100~h~\rm{km}~\rm{sec}^{-1}~\rm{Mpc}^{-1}$
($h\approx 0.72\pm 0.07$ \cite{h}).
\vspace{.25cm}
\item[{\bf ii.}] The fraction $\Omega=\rho/\rho_{c}$,
where $\rho_{c}$ is the critical density corresponding
to a flat universe. From (\ref{eq:friedmann}),
$\rho_{c}=3H^{2}/8\pi G$ and $\Omega=1+k/a^{2}H^{2}$.
$\Omega=1$, $>1$ or $<1$ corresponds to flat, closed or
open universe. Assuming inflation \index{inflation} (see
below), the present value of $\Omega$ must be
$\Omega_{0}=1$ in accord with the recent DASI observations
which yield \cite{dasi} $\Omega_{0}=1\pm 0.04$. The low
deuterium abundance measurements \cite{deuterium} give
$\Omega_{B}h^2\approx 0.020\pm 0.001$, where $\Omega_{B}$
is the baryonic contribution to $\Omega_0$. This result
implies that $\Omega_{B}\approx 0.039\pm 0.077$. The total
contribution $\Omega_M$ of matter to $\Omega_0$ can then be
determined from the measurements \cite{cluster} of the
baryon-to-matter ratio in clusters. It is found that
$\Omega_M\approx 1/3$, which shows that most of the matter
in the universe is non-baryonic, i.e., dark matter.
Moreover, we see that about $2/3$ of the energy density of
the universe is not even in the form of matter and we call
it dark energy.
\vspace{.25cm}
\item[{\bf iii.}] The deceleration parameter
\begin{equation}
q=-\frac{(\ddot{a}/\dot{a})}{(\dot{a}/a)}
=\frac{\rho+3p}{2\rho_{c}}~\cdot
\label{eq:decel}
\end{equation}
Measurements of type Ia supernovae \cite{lambda} indicate
that the universe is speeding up ($q_0<0$). This requires
that, at present, $p<0$ as can be seen from
(\ref{eq:decel}). Negative pressure can only be
attributed to the dark energy since matter is pressureless.
Equation (\ref{eq:decel}) gives
$q_0=(\Omega_0+3w_X\Omega_X)/2$, where
$\Omega_X=\rho_X/\rho_c$ and $w_X=p_X/\rho_X$ with
$\rho_X$ and $p_X$ being the dark energy density and
pressure. Observations prefer $w_X=-1$, with a 95\%
confidence limit $w_X<-0.6$ \cite{w}. Thus, dark energy can
be interpreted as something close to a non-zero cosmological
constant (see below).
\end{list}

\subsection{Particle Horizon \index{horizon}}
\label{subsec:parhor}

\par
Light travels only a finite distance from the time of big bang
\index{big bang} ($t=0$) until some cosmic time $t$. From
(\ref{eq:rw}), we find that the propagation of light along the
radial direction is described by $a(t)dr=dt$. The particle
horizon, \index{horizon} which is the `instantaneous' distance
at $t$ travelled by light since $t=0$, is then
\begin{equation}
d_{H}(t)=a(t)\int_{0}^{t}\frac{dt^{\prime}}
{a(t^{\prime})}~\cdot
\label{eq:hor}
\end{equation}
The particle horizon \index{horizon} is an important
notion since it coincides with the size of the universe already
seen at time $t$ or, equivalently, with the distance at which
causal contact has been established at $t$. Equations
(\ref{eq:expan}) and (\ref{eq:hor}) give
\begin{equation}
d_{H}(t)=\frac{3\gamma}{3\gamma-2}t~,
~\gamma\neq 2/3~.
\label{eq:hort}
\end{equation}
Also,
\begin{equation}
H(t)=\frac{2}{3\gamma}t^{-1},
~d_{H}(t)=\frac{2}{3\gamma-2}H^{-1}(t)~.
\label{eq:hubblet}
\end{equation}
For `matter' (`radiation'), these formulae become $d_{H}(t)
=2H^{-1}(t)=3t$ ($d_{H}(t)=H^{-1}(t)=2t$). Assuming
matter dominance, the present particle horizon
\index{horizon} (cosmic time) is $d_{H}(t_{0})=
2H_{0}^{-1}\approx 6,000~h^{-1}~{\rm{Mpc}}$
($t_{0}=2H_{0}^{-1}/3\approx 6.5\times 10^{9}
~h^{-1}~{\rm{years}}$). The present
$\rho_{c}=3H_{0}^{2}/8\pi G\approx 1.9\times
10^{-29}~h^2~{\rm{gm/cm^{3}}}$.

\subsection{Brief History of the Early Universe}
\label{subsec:history}

\par
We will now briefly describe the early stages of the universe
evolution according to GUTs \index{GUT} \cite{ggps}. We take
a GUT \index{GUT} based on the gauge group $G$ ($=SU(5)$,
$SO(10)$, $SU(3)^{3}$, ...) with or without SUSY.
\index{SUSY} At a superheavy scale
$M_{X}\sim 10^{16}~{\rm{GeV}}$ (the GUT \index{GUT} mass
scale), $G$ breaks to the standard model gauge group
$G_{S}=SU(3)_{c}\times SU(2)_{L}\times U(1)_{Y}$ by the
vev of an appropriate Higgs field $\phi$. (For simplicity, we
consider that this breaking occurs in one step.) $G_{S}$ is,
subsequently, broken to $SU(3)_{c}\times U(1)_{em}$ at the
electroweak scale $M_{W}$.

\par
GUTs \index{GUT} together with the SBB \index{SBB} model
provide a suitable framework for discussing the early history
of the universe for cosmic times
$\stackrel{_{>}}{_{\sim }} 10^{-44}~{\rm{sec}}$.
They predict that the universe, as it expands and cools
after the big bang, \index{big bang} undergoes \cite{kl} a
series of phase transitions during which the gauge symmetry is
gradually reduced and several important phenomena take place.

\par
After the big bang, \index{big bang} $G$ was unbroken and
the universe was filled with a hot `soup' of massless particles
which included not only photons, quarks, leptons and gluons but
also the weak gauge boson $W^{\pm}$, $Z^{0}$, the GUT
\index{GUT} gauge bosons $X$, $Y$, ... and several Higgs
bosons. (In the SUSY \index{SUSY} case, all the SUSY
\index{SUSY} partners of these particles were also present.)
At cosmic time $t\sim 10^{-37}~{\rm{sec}}$ corresponding to
temperature $T\sim 10^{16}~{\rm{GeV}}$, $G$ broke down to
$G_{S}$ and the $X$, $Y$, ... gauge bosons together with some
Higgs bosons acquired superheavy masses of order $M_{X}$.
Their out-of-equilibrium decay could, in principle, produce
\cite{dimopoulos,bau} the observed BAU \index{BAU} (with
the reservation at the end of Sect.\ref{subsec:sphaleron}).
Important ingredients are the violation of baryon number,
which is inherent in GUTs, \index{GUT} and C and CP
violation. This is the second (potential) success of SBB.
\index{SBB}

\par
During the GUT \index{GUT} phase transition, topologically
stable extended objects \cite{kibble} such as monopoles
\index{monopole} \cite{monopole}, cosmic strings
\index{cosmic strings} \cite{string} or domain walls
\index{domain walls} \cite{wall} can also be produced.
Monopoles, \index{monopole} which exist in most GUTs,
\index{GUT} can lead into cosmological problems
\cite{preskill} which are, however, avoided by inflation
\cite{guth,lindebook} \index{inflation} (see
Sects.\ref{subsec:monopole} and \ref{subsec:infmono}). This
is a period of an exponentially fast expansion of the universe
which can occur during some GUT \index{GUT} phase transition.
Cosmic strings \index{cosmic strings} can contribute
\cite{zel} to the primordial density perturbations
\index{density perturbations} necessary for structure
formation \index{structure formation} \cite{structure} in
the universe whereas domain walls \index{domain walls} are
\cite{wall} absolutely catastrophic and GUTs \index{GUT}
should be constructed so that they avoid them (see e.g.,
\cite{axion}) or inflation \index{inflation} should be used
to remove them from the scene.

\par
At $t\sim 10^{-10}~{\rm{sec}}$ or
$T\sim 100~{\rm{GeV}}$,
the electroweak transition takes place and $G_{S}$ breaks
to $SU(3)_{c}\times U(1)_{em}$. $W^{\pm}$, $Z^{0}$
and the electroweak Higgs fields acquire masses
$\sim M_{W}$. Subsequently, at
$t\sim 10^{-4}~{\rm{sec}}$ or $T\sim 1~{\rm{GeV}}$,
color confinement sets in and the quarks get bounded
forming hadrons.

\par
The direct involvement of particle physics essentially ends
here since most of the subsequent phenomena fall into the
realm of other branches. We will, however, sketch some of
them since they are crucial for understanding the earlier
stages of the universe evolution where their origin lies.

\par
At $t\approx 180~{\rm{sec}}$ ($T\approx 1~{\rm{MeV}}$),
nucleosynthesis \index{nucleosynthesis} takes place, i.e.,
protons and neutrons form nuclei. The abundance of light
elements ($D$, $^{3}He$, $^{4}He$ and $^{7}Li$) depends
\cite{peebles} crucially on the number of light particles
(with mass $\stackrel{_{<}}{_{\sim }} 1~{\rm{MeV}}$),
i.e., the number of light neutrinos, $N_{\nu}$, and
$\Omega_{B}h^{2}$. Agreement with observations
\cite{deuterium} is achieved for $N_{\nu}=3$ and
$\Omega_{B}h^{2}\approx 0.020$. This is the third success
of SBB \index{SBB} cosmology. Much later, at the so-called
`equidensity' point, \index{`equidensity' point}
$t_{\rm{eq}}\approx 5\times 10^4~{\rm{years}}$, matter
dominates over radiation.

\par
At cosmic time $t\approx 200,000~h^{-1} {\rm{years}}$
($T\approx 3,000~{\rm{K}}$), we have the `decoupling'
\index{decoupling} of matter and radiation and the
`recombination' of atoms. After this, radiation
evolves as an independent (not interacting) component
of the universe and is detected today as CMBR \index{CMBR}
with temperature $T_{0}\approx 2.73~{\rm{K}}$. The
existence of this radiation is the fourth success of
the SBB \index{SBB} model. Finally, structure formation
\index{structure formation} \cite{structure} in the
universe starts at
$t\approx 2\times 10^{8}~{\rm{years}}$.

\section{Shortcomings of Big Bang \index{big bang}}
\label{sec:short}

The SBB \index{SBB} cosmological model has been very successful
in explaining, among other things, the Hubble expansion of the
universe, the existence of the CMBR \index{CMBR} and the
abundances of the light elements which were formed during
primordial nucleosynthesis. \index{nucleosynthesis} Despite its
great successes, this model had a number of long-standing
shortcomings which we will now summarize:

\subsection{Horizon \index{horizon} Problem}
\label{subsec:horizon}

The CMBR, \index{CMBR} which we receive now, was emitted at
the time of `decoupling' \index{decoupling} of matter and
radiation when the cosmic temperature was $T_d \approx
3,000~\rm{K}$. The decoupling \index{decoupling} time,
$t_d$, can be calculated from
\begin{equation}
\frac {T_0}{T_d} =
\frac {2.73~\rm{K}}{3,000~\rm{K}} =
\frac {a (t_d)}{a(t_0)} =
\left(\frac {t_d}{t_0}\right)^{2/3}\cdot
\label{eq:dec}
\end{equation}
It turns out that $t_d \approx 200,000~h^{-1}$ years.

\par
The distance over which the CMBR \index{CMBR} has
travelled since its emission is
\begin{equation}
a(t_0) \int^{t_{0}} _{t_{d}}
\frac {dt^\prime}{a(t^\prime)} = 3t_0
\left[1 - \left(\frac {t_d}{t_0}\right)^{2/3}\right]
\approx 3t_0 \approx 6,000~h^{-1}~\rm{Mpc}~,
\label{eq:lss}
\end{equation}
which essentially coincides with the present particle horizon
\index{horizon} size. A sphere around us with radius
equal to this distance is called the `last scattering surface'
\index{last scattering surface} since the CMBR \index{CMBR}
observed now has been emitted from it. The particle horizon
\index{horizon} size at $t_d$ was $2H^{-1}(t_d)=3t_d \approx
0.168~h^{-1}~\rm{Mpc}$ and expanded until now to become
equal to $0.168~h^{-1} (a(t_0)/a(t_d))~{\rm{Mpc}}\approx
184~h^{-1}$ Mpc. The angle subtended by this `decoupling'
\index{decoupling} horizon \index{horizon} at present is
$\theta_{d} \approx 184/6,000 \approx 0.03~\rm{rads}
\approx 2^{o}$. Thus, the sky splits into $4 \pi/(0.03)^2
\approx 14,000$ patches which never communicated causally
before sending light to us. The question then arises how
come the temperature of the black body radiation from all
these patches is so accurately tuned as the results of COBE
\index{COBE} \cite{cobe} require.

\subsection{Flatness Problem \index{flatness problem}}
\label{subsec:flatness}

The present energy density of the universe has been observed
\cite{dasi} to be very close to its critical energy density
corresponding to a flat universe ($\Omega_{0}=1\pm 0.04$).
Equation (\ref{eq:friedmann}) implies that
$(\rho-\rho_c)/\rho_c=3(8\pi G\rho_c)^{-1}(k/a^2)$
is proportional to $a$, for matter dominated universe.
Thus, in the early universe, we have
$|(\rho-\rho_c)/\rho_c|\ll 1$ and the question  arises
why the initial energy density of the universe was so finely
tuned to be equal to its critical value.

\subsection{Magnetic Monopole \index{monopole} Problem}
\label {subsec:monopole}

This problem arises only if we combine the SBB \index{SBB}
model with GUTs \index{GUT} \cite{ggps} which predict the
existence of magnetic monopoles. \index{monopole} As already
indicated, according to GUTs, \index{GUT} the universe underwent
\cite{kl} a phase transition during which the GUT \index{GUT}
gauge symmetry group, $G$, broke to $G_{S}$. This breaking was
due to the fact that, at a critical temperature $T_c$, an
appropriate Higgs field, $\phi$, developed a non-zero vev.
Assuming that this phase transition was a second order one, we
have $\langle\phi\rangle(T)\approx\langle\phi\rangle(T=0)
(1-T^2/T^2_c)^{1/2}$, $m_H (T)\approx\lambda\langle\phi
\rangle(T)$, for the temperature dependent vev and mass of the
Higgs field respectively at $T\leq T_c$ ($\lambda$ is an
appropriate Higgs coupling constant).

\par
The GUT \index{GUT} phase transition produces monopoles
\index{monopole} \cite{monopole} which are localized deviations
from the vacuum with radius $\sim M_X^{-1}$, mass $m_M\sim M_X/
\alpha_G$ and $\phi =0$ at their center ($\alpha_G=g^2_{G}
/4\pi$ with $g_G$ being the GUT \index{GUT} gauge coupling
constant). The vev of the Higgs field on a sphere, $S^2$, with
radius $\gg M_{X}^{-1}$ around the monopole \index{monopole}
lies on the vacuum manifold $G/G_S$ and we, thus, obtain a
mapping: $S^2\longrightarrow G/G_S$. If this mapping is
homotopically non-trivial the topological stability of the
monopole \index{monopole} is guaranteed.

\par
Monopoles \index{monopole} can be produced when the
fluctuations of $\phi$ over $\phi=0$ between the vacua at
$\pm\langle\phi\rangle(T)$ cease to be frequent. This
occurs when the free energy needed for $\phi$ to fluctuate
from $\langle\phi\rangle(T)$ to zero in
a region of radius equal to the Higgs correlation length
$\xi(T)=m^{-1}_H (T)$ exceeds $T$. This condition
reads $(4\pi/3)\xi^3\Delta V\stackrel{_{>}}
{_{\sim }}T$, where $\Delta V\sim\lambda^2\langle
\phi\rangle^4$ is the difference in free energy density
between $\phi=0$ and $\phi=\langle\phi\rangle(T)$.
The Ginzburg temperature \index{Ginzburg temperature}
\cite{ginzburg}, $T_G$, corresponds to the saturation
of this inequality. So, at
$T\stackrel{_{<}}{_{\sim }}T_G$, the fluctuations
over $\phi=0$ stop and $\langle\phi\rangle$ settles
on $G/G_S$. At $T_G$, the universe splits into regions
of size $\xi_G\sim(\lambda^2 T_c)^{-1}$,
the Higgs correlation length at $T_G$, with $\phi$
being more or less aligned in each region. Monopoles
\index{monopole} are produced at the corners where such
regions meet (Kibble mechanism \index{Kibble mechanism}
\cite{kibble}) and their number density is estimated to
be $n_M\sim{\rm{p}}\xi_{G}^{-3}\sim{\rm{p}}
\lambda^{6}T_{c}^{3}$, where $\rm{p}\sim\rm{1/10}$
is a geometric factor. The `relative' monopole
\index{monopole} number density then turns out to be
$r_M=n_M/T^3\sim\rm{10^{-6}}$. We can derive a lower
bound on $r_M$ by employing causality. The Higgs field
$\phi$ cannot be correlated at distances bigger than the
particle horizon \index{horizon} size, $2t_G$, at $T_G$.
This gives the causality bound
\begin{equation}
n_M  \stackrel{_{>}}{_{\sim }}\frac {\rm{p}}
{\frac{4 \pi}{3}(2t_G)^3}~,
\label{eq:causal}
\end{equation}
which implies that $r_M\stackrel{_{>}}{_{\sim }}
\rm{10^{-10}}$.

\par
The subsequent evolution of monopoles, \index{monopole}
after $T_G$, is governed by \cite{preskill}
\begin{equation}
\frac {dn_M}{dt} =
- D n_{M}^{2} - 3 \frac {\dot{a}}{a} n_{M}~,
\label{eq:evol}
\end{equation}
where the first term in the rhs (with $D$ being an
appropriate constant) describes the dilution of monopoles
\index{monopole} by their annihilation with antimonopoles,
while the second term corresponds to their dilution by
Hubble expansion. The monopole-antimonopole \index{monopole}
annihilation proceeds as follows. Monopoles \index{monopole}
diffuse towards antimonopoles in the plasma of charged
particles, capture each other in Bohr orbits and eventually
annihilate. The annihilation is effective provided that the
mean free path of monopoles \index{monopole} in the plasma
does not exceed their capture distance. This holds for
$T\stackrel{_{>}}{_{\sim }}\rm{10^{12}~GeV}$. The
overall result is that, if the initial relative monopole
\index{monopole} density $r_{M,\rm{in}}
\stackrel{_{>}}{_{\sim }}\rm{10^{-9}}$
($\stackrel{_{<}}{_{\sim }}\rm{10^{-9}}$), the final one
$r_{M,\rm{fin}}\sim 10^{-9}$ ($\sim r_{M,\rm{in}}$).
This combined with the causality bound yields
$r_{M,\rm{fin}} \stackrel{_{>}}{_{\sim }}
\rm{10^{-10}}$. However, the requirement that monopoles
\index{monopole} do not dominate the energy density of
the universe at nucleosynthesis \index{nucleosynthesis} gives
\begin{equation}
r_M (T \approx 1~\rm{MeV}) \stackrel{_{<}}
{_{\sim }}\rm{10^{-19}},
\label{eq:nucleo}
\end{equation}
and we obtain a clear discrepancy of about ten orders
of magnitude.

\subsection{Density Perturbations
\index{density perturbations}}
\label{subsec:fluct}

For structure formation \index{structure formation}
\cite{structure} in the universe, we need a primordial
density perturbation, \index{density perturbations}
$\delta \rho/ \rho$, at all length scales with a nearly
flat spectrum \cite{hz}. We also need an explanation
of the temperature fluctuations
\index{temperature fluctuations} of the CMBR \index{CMBR}
observed by COBE \index{COBE} \cite{cobe} at angles
$\theta\stackrel{_{>}}{_{\sim }}\theta_d\approx 2^{o}$
which violate causality (see Sect.\ref{subsec:horizon}).

\par
Let us expand $\delta\rho/\rho$ in plane waves
\begin{equation}
\frac {\delta \rho} {\rho} (\bar{r},t) =
\int d^3 k\delta_{\bar{k}}(t)e^{i\bar{k} \bar{r}},
\label{eq:plane}
\end{equation}
where $\bar{r}$ is a `comoving' vector in 3-space and
$\bar{k}$ is the `comoving' wave vector with
$k=|\bar{k}|$ being the `comoving' wave number
($\lambda=2 \pi/k$ is the `comoving' wave length and
the physical wave length is $ \lambda _{\rm{phys}}=
a(t) \lambda$). For $\lambda_{\rm{phys}} \leq H^{-1}$,
the time evolution of $\delta_{\bar{k}}$ is described by
the Newtonian equation \index{Newtonian equation}
\begin{equation}
\ddot{\delta}_{\bar{k}} + 2 H \dot {\delta}_{\bar{k}}
+ \frac {v_{s}^{2}k^2}{a^2}\delta_{\bar{k}}=
4 \pi G \rho \delta_{\bar{k}}~,
\label{eq:newton}
\end{equation}
where the second term in the left hand side (lhs) comes
from Hubble expansion and the third is the `pressure
term' ($v_s$ is the velocity of sound given by
$v^{2}_{s}=dp/d\rho$). The rhs corresponds to the
gravitational attraction.

\par
For the moment, put $H$=0 (static universe). There exists
then a characteristic wave number $k_J$, the Jeans wave
number, given by $k^{2}_{J}=4\pi G a^{2}\rho/v^{2}_{s}$
and having the following property. For $k>k_J$,  pressure
dominates over gravitational attraction and the density
perturbations \index{density perturbations} just oscillate,
whereas, for $k<k_J$,
attraction dominates and the perturbations grow
exponentially. In particular, for `matter', $v_s=0$ and
all scales are Jeans unstable with
\begin{equation}
\delta_{\bar{k}} \propto {\rm{exp}}(t/\tau)~,~\tau=
(4 \pi G \rho)^{-1/2}.
\label{eq:jeans}
\end{equation}

\par
Now let us take $H\neq 0$. Since the cosmological expansion
pulls the particles apart, we get a smaller growth:
\begin{equation}
\delta_{\bar{k}}\propto a(t) \propto t^{2/3},
\label{eq:growth}
\end{equation}
in the matter dominated case. For `radiation' ($p \neq 0$),
we get essentially no growth of the density perturbations.
\index{density perturbations}
This means that, in order to have structure formation
\index{structure formation} in the universe, which requires
$\delta \rho/\rho \sim 1$, we must have
\begin{equation}
(\frac {\delta \rho} {\rho})_{\rm{eq}}\sim
4\times 10^{-5} (\Omega_M h^2)^{-1},
\label{eq:equi}
\end{equation}
at the `equidensity' point, \index{`equidensity' point}
since the available growth factor for perturbations is
given by $a_0/a_{\rm{eq}}\sim 2.5\times 10^4\Omega_M
h^2$. The question then is where these primordial
density perturbations \index{density perturbations}
originate from.

\section{Inflation \index{inflation}}
\label{sec:inflation}

Inflation \cite{guth,lindebook} \index{inflation} is
an idea which solves
simultaneously all four cosmological puzzles and can be
summarized as follows. Suppose there is a real scalar
field $\phi$ (the inflaton) \index{inflaton} with
(symmetric) potential energy density $V(\phi)$ which
is quite flat near $\phi=0$ and has minima at $\phi=
\pm\langle\phi\rangle$ with
$V(\pm\langle\phi\rangle)=0$.
At high enough $T$'s, $\phi=0$ in the universe due to
the temperature corrections to $V(\phi)$. As $T$ drops,
the effective potential approaches the $T$=0
potential but a little potential barrier separating
the local minimum at $\phi=0$ and the vacua at
$\phi=\pm\langle\phi\rangle$ still remains.
At some point, $\phi$ tunnels out to
$\phi_1\ll\langle\phi\rangle$ and a bubble with
$\phi=\phi_1$ is created in the universe. The field
then rolls over to the minimum of $V(\phi)$ very slowly
(due to the flatness of the potential). During this slow
roll-over, the energy density
$\rho\approx V(\phi=0)\equiv V_0$ remains essentially
constant for quite some time. The Lagrangian density
\begin{equation}
L=\frac{1}{2}\partial_{\mu}\phi\partial^{\mu}\phi
-V(\phi)
\label{eq:lagrange}
\end{equation}
gives the energy momentum tensor
\begin{equation}
T_{\mu}^{~\nu}=-\partial_\mu\phi\partial^\nu \phi+
\delta_{\mu}^{~\nu}\left(\frac{1}{2}
\partial_\lambda\phi\partial^\lambda\phi-
V (\phi)\right),
\label{eq:energymom}
\end{equation}
which during the slow roll-over takes the form
$T_{\mu}^{~\nu}\approx -V_{0}~\delta_{\mu}^{~\nu}$.
This means that $\rho \approx -p\approx V_0$, i.e.,
the pressure is negative and equal in magnitude with the
energy density, which is consistent with
(\ref{eq:continuity}). As we will see, $a(t)$ grows
fast and the `curvature term', $k/a^2$, in
(\ref{eq:friedmann}) diminishes. We thus get
\begin{equation}
H^2\equiv\left(\frac{\dot{a}}{a}\right)^2=
\frac{8\pi G}{3}V_0~,
\label{eq:inf}
\end{equation}
which gives $a(t)\propto e^{Ht}$, $H^2=(8 \pi G/3)V_0=
{\rm constant}$. So the bubble expands exponentially for some
time and $a(t)$ grows by a factor
\begin{equation}
\frac {a(t_f)}{a(t_i)}={\rm{exp}}H(t_f-t_i)
\equiv{\rm{exp}}H\tau~,
\label{eq:efold}
\end{equation}
between an initial ($t_i$) and a final ($t_f$) time.

\par
The inflationary scenario just described, known as `new'
\cite{new} inflation \index{inflation} (with the
inflaton \index{inflaton} starting from zero), is not
the only realization of the idea of inflation.
\index{inflation} Another possibility is to consider
the universe as it emerges at $t_{P}$. We can imagine
a region of size $\ell_{P}\sim M_{P}^{-1}$ (the
Planck length) where the inflaton \index{inflaton}
acquires a large and almost uniform value and carries
negligible kinetic energy. Under certain circumstances,
this region can inflate (exponentially expand) as $\phi$
rolls down towards the vacuum. This type of inflation
\index{inflation} with the inflaton \index{inflaton}
starting from large values is known as `chaotic'
\cite{chaotic} inflation. \index{inflation}

\par
We will now show that, with an adequate number of e-foldings,
$N=H\tau$, the first three cosmological puzzles are easily
resolved (we leave the question of density perturbations
\index{density perturbations} for later).

\subsection{Resolution of the Horizon \index{horizon} Problem}
\label{subsec:infhor}

The particle horizon \index{horizon} during inflation
\index{inflation}
\begin{equation}
d(t)=e^{Ht}\int^t_{t_{i}}
\frac{d t^\prime}{e^{Ht^\prime}}
\approx H^{-1}{\rm{exp}}H(t-t_i)~,
\label{eq:horizon}
\end{equation}
for $t-t_i \gg H^{-1}$, grows as fast as $a(t)$. At the
end of inflation \index{inflation} ($t=t_f$),
$d(t_f)\approx H^{-1}{\rm{exp}} H \tau$ and $\phi$
starts oscillating about the minimum of the potential at
$\phi=\langle\phi\rangle$. It finally decays and `reheats'
\cite{reheat} the universe at a temperature
$T_r\sim 10^9~{\rm{GeV}}$ \cite{gravitino}. The universe
then returns to normal big bang \index{big bang} cosmology.
The horizon \index{horizon}
$d(t_{f})$ is stretched during the $\phi$-oscillations by
a factor $\sim 10^9$ depending on details and between $T_r$
and the present by a factor $T_r/T_0$. So it finally becomes
equal to $H^{-1}e^{H\tau}10^9(T_r/T_0)$, which should
exceed $2H_{0}^{-1}$ in order to solve the horizon
\index{horizon} problem. Taking $V_0\approx M_{X}^{4}$,
$M_{X}\sim 10^{16}~{\rm GeV}$, we see that, with
$N=H\tau\stackrel{_{>}}{_{\sim }}55$, the horizon
\index{horizon} problem is evaded.

\subsection{Resolution of the Flatness Problem
\index{flatness problem}}
\label{subsec:infflat}

The `curvature term' of the Friedmann equation,
\index{Friedmann equation} at present, is given by
\begin{equation}
\frac{k}{a^2}\approx\left(\frac{k}{a^2}\right)_{bi}
e^{-2H\tau}~10^{-18}\left(\frac {10^{-13}~{\rm{GeV}}}
{10^9~{\rm{ GeV}}}\right)^2,
\label{eq:curvature}
\end{equation}
where the terms in the rhs correspond to the `curvature
term' before inflation, \index{inflation} and its growth
factors during inflation, \index{inflation} during
$\phi$-oscillations and after `reheating'
\index{reheating} respectively. Assuming
$(k/a^2)_{bi}\sim(8\pi G/3)\rho\sim H ^2$
($\rho\approx V_0$), we obtain
$\Omega_0-1=k/a_{0}^{2}H_{0}^{2}\sim
10^{48}~e^{-2H \tau}$ which is $\ll 1$, for
$H\tau\gg 55$. Strong inflation \index{inflation}
implies that the present universe is flat with a great
accuracy.

\subsection{Resolution of the Monopole \index{monopole}
Problem}
\label{subsec:infmono}

For $N\stackrel{_{>}}{_{\sim }} 55$, the monopoles
\index{monopole} are diluted by at least 70 orders of
magnitude and  become irrelevant. Also, since
$T_r \ll m_M$, there is no monopole \index{monopole}
production after `reheating'. \index{reheating} Extinction
of monopoles \index{monopole} may also be achieved by
non-inflationary
mechanisms such as magnetic confinement \cite{fate}. For
models leading to a possibly measurable monopole
\index{monopole} density see e.g.,
\cite{thermal,trinification}.

\section{Detailed Analysis of Inflation \index{inflation}}
\label{sec:detail}

The Hubble parameter is not exactly constant during
inflation \index{inflation} as we, naively, assumed so far.
It actually depends on the value of $\phi$:
\begin{equation}
H^{2}(\phi)=\frac{8\pi G}{3}V(\phi)~.
\label{eq:hubble}
\end{equation}
To find the evolution equation for $\phi$ during inflation,
\index{inflation} we vary the action
\begin{equation}
\int\sqrt{-{\rm{det}}(g)}~d^{4}x\left(\frac{1}{2}
\partial_ {\mu}\phi\partial^{\mu}\phi-V(\phi)+
M(\phi)\right),
\label{eq:action}
\end{equation}
where $g$ is the metric tensor and $M(\phi)$ represents
the coupling of $\phi$ to `light' matter causing its decay.
We find
\begin{equation}
\ddot{\phi}+3H\dot{\phi}+\Gamma_{\phi}\dot{\phi}+
V^{\prime}(\phi)=0~,
\label{eq:evolution}
\end{equation}
where the prime denotes derivation with respect to $\phi$
and $\Gamma_{\phi}$ is the decay width \cite{width} of
the inflaton. \index{inflaton} Assume, for the moment,
that the decay time
of $\phi$, $t_d=\Gamma_{\phi}^{-1}$, is much greater
than $H^{-1}$, the expansion time for inflation.
\index{inflation} Then the term
$\Gamma_{\phi}\dot{\phi}$ can be ignored and
(\ref{eq:evolution}) becomes
\begin{equation}
\ddot{\phi}+3H\dot{\phi}+V^{\prime}(\phi)=0~.
\label{eq:reduce}
\end{equation}
Inflation \index{inflation} is by definition the
situation where
$\ddot{\phi}$ is subdominant to the `friction term'
$3H\dot{\phi}$ (and the kinetic energy density is
subdominant to the potential one). Equation
(\ref{eq:reduce}) then reduces to the inflationary
equation \cite{slowroll}
\begin{equation}
3H\dot{\phi}=-V^{\prime}(\phi)~,
\label{eq:infeq}
\end{equation}
which gives
\begin{equation}
\ddot{\phi}=
-\frac{V^{\prime\prime}(\phi)\dot{\phi}}
{3H(\phi)}+\frac{V^{\prime}(\phi)}
{3H^{2}(\phi)}H^\prime(\phi)\dot{\phi}~.
\label{eq:phidd}
\end{equation}
Comparing the two terms in the rhs of this equation with
the `friction term' in (\ref{eq:reduce}), we get the
conditions for inflation \index{inflation} (slow roll
conditions):
\begin{equation}
|\eta|\equiv\frac{M_{P}^{2}}{8 \pi}\bigg|
\frac{V^{\prime\prime}(\phi)}
{V(\phi)}\bigg|\leq 1~,
~\epsilon \equiv \frac {M_{P}^{2}}{16 \pi}
\left(\frac {V^{\prime}(\phi)}
{V(\phi)}\right)^{2} \leq 1~.
\label{eq:src}
\end{equation}
The end of the slow roll-over occurs when either of these
inequalities is saturated. If $\phi_f$ is the value of
$\phi$ at the end of inflation, \index{inflation} then
$t_f\sim H^{-1}(\phi_f)$.

\par
The number of e-foldings during inflation \index{inflation}
can be calculated as follows:
\begin{equation}
N(\phi_{i}\rightarrow \phi_{f})\equiv\ln
\left(\frac{a(t_{f})}
{a(t_{i})}\right)=\int^{t_{f}} _{t_{i}} Hdt=
\int^{\phi_{f}}_{\phi_{i}}
\frac{H (\phi)}{\dot{\phi}}d\phi=-
\int^{\phi_{f}}_{\phi_{i}}
\frac {3 H^2 (\phi) d \phi}
{V^{\prime}(\phi)},
\label{eq:nefolds}
\end{equation}
where (\ref{eq:efold}), (\ref{eq:infeq}) and the
definition of $H=\dot{a}/a$ were used. For simplicity,
we can shift the field $\phi$ so that the global minimum of
the potential is displaced at $\phi$ = 0. Then, if
$V(\phi) = \lambda \phi^{\nu}$ during inflation,
\index{inflation} we have
\begin{equation}
N(\phi_{i} \rightarrow \phi_{f}) =
- \int^{\phi_{f}}_{\phi_{i}} \frac
{3H^2(\phi)d\phi}{V^{\prime}(\phi)} =
- 8 \pi G \int^{\phi_{f}}_{\phi_{i}} \frac
{V(\phi)d\phi}{V^{\prime}(\phi)}=\frac {4 \pi G}{\nu}
(\phi^{2}_{i}-\phi^{2}_{f})~.
\label{eq:expefold}
\end{equation}
Assuming that $\phi_{i} \gg \phi_{f}$, this reduces to
$N(\phi)\approx(4 \pi G/\nu)\phi^2$.

\section{Coherent Oscillations of the Inflaton
\index{inflaton}}
\label{sec:osci}

After the end of inflation \index{inflation} at $t_f$, the
term $\ddot{\phi}$
takes over in (\ref{eq:reduce}) and $\phi$ starts
performing coherent damped oscillations about the global
minimum of the potential. The rate of energy density loss, due
to `friction', is given by
\begin{equation}
\dot{\rho}=\frac{d}{dt}\left(\frac{1}{2}
\dot{\phi}^2+V(\phi)\right)=-3H\dot{\phi}^2=
-3H(\rho+p)~,
\label{eq:damp}
\end{equation}
where $\rho=\dot{\phi}^2/2+V(\phi)$ and
$p=\dot{\phi}^2/2-V(\phi)$. Averaging $p$ over one
oscillation of $\phi$ and writing $\rho+p=\gamma\rho$, we
get $\rho\propto a^{-3 \gamma}$ and $a(t)\propto
t^{2/3\gamma}$ (see Sect.\ref{subsec:friedmann}).

\par
The number $\gamma$ can be written as (assuming a symmetric
potential)
\begin{equation}
\gamma=\frac{\int^{T}_{0}\dot{\phi}^{2}dt}
{\int^{T}_{0} \rho dt}=
\frac{\int^{\phi_{{\rm{max}}}}_{0}
\dot{\phi}d\phi}
{\int^{\phi_{{\rm{max}}}}_{0}
(\rho/\dot{\phi})d\phi}~,
\label{eq:gamma}
\end{equation}
where $T$ and $\phi_{{\rm{max}}}$ are the period and
the amplitude of the oscillation. From
$\rho=\dot{\phi}^2/2+V(\phi)=V_{{\rm{max}}}$, where
$V_{\rm max}$ is the maximal potential energy density, we
obtain $\dot{\phi}=\sqrt{2(V_{{\rm{max}}}-V(\phi))}$.
Substituting this in (\ref{eq:gamma}) we get
\cite{oscillation}
\begin{equation}
\gamma=\frac {2\int^{\phi_{{\rm{max}}}}_{0}
(1-V/V_{{\rm{max}}})^{1/2}
d\phi}{\int^{\phi_{{\rm{max}}}}_{0}
(1-V/V_{{\rm{max}}})^{-1/2}d\phi}~\cdot
\label{eq:gammafinal}
\end{equation}
For $V(\phi)=\lambda\phi^{\nu}$, we find $\gamma=
2\nu/(\nu+2)$ and, thus, $\rho\propto
a^{-6\nu/(\nu+2)}$ and $a(t)\propto t^{(\nu+2)/3\nu}$.
For $\nu=2$, in particular, $\gamma=1$,
$\rho\propto a^{-3}$, $a(t)\propto t^{2/3}$ and
$\phi$ behaves like pressureless matter.
This is not unexpected since a coherent oscillating massive
free field corresponds to a distribution of static massive
particles. For $\nu$=4, we obtain $\gamma=4/3$,
$\rho\propto a^{-4}$, $a(t)\propto t^{1/2}$ and the
system resembles radiation. For $\nu = 6$, one has
$\gamma=3/2$, $\rho\propto a^{-9/2}$,
$a(t)\propto t^{4/9}$ and the expansion is slower (the
pressure is higher) than in radiation.

\section{Decay of the Inflaton \index{inflaton}}
\label{sec:decay}

Reintroducing the `decay term'
$\Gamma_{\phi} \dot{\phi}$,
(\ref{eq:evolution}) can be written as
\begin{equation}
\dot{\rho}=
\frac{d}{dt}\left(\frac{1}{2}\dot{\phi}^2+
V(\phi)\right)=-(3H+\Gamma_\phi)\dot{\phi}^2,
\label{eq:decay}
\end{equation}
which is solved \cite{reheat,oscillation} by
\begin{equation}
\rho(t)=\rho_{f}
\left(\frac{a(t)}{a(t_{f})}\right)^{-3 \gamma}
{\rm{exp}}[-\gamma \Gamma_{\phi}(t-t_f)]~,
\label{eq:rho}
\end{equation}
where $\rho_f$ is the energy density at $t_f$. The second
and third factors in the rhs of this equation represent the
dilution of the field energy due to the expansion of the
universe and the decay of $\phi$ to `light' particles
respectively.

\par
All pre-existing radiation (known as `old radiation')
was diluted by inflation, \index{inflation} so the only
radiation present is the one produced by the decay of
$\phi$ and is known as
`new radiation'. Its energy density satisfies
\cite{reheat,oscillation} the equation
\begin{equation}
\dot{\rho}_{r}=-4H \rho_{r}+
\gamma\Gamma_{\phi}\rho~,
\label{eq:newrad}
\end{equation}
where the first term in the rhs represents the dilution of
radiation due to the cosmological expansion while the
second one is the energy density transfer from $\phi$ to
radiation. Taking $\rho_{r}(t_f)$=0, this equation gives
\cite{reheat,oscillation}
\begin{equation}
\rho_{r}(t) = \rho_{f}\left(\frac {a(t)}
{a(t_{f})}\right)^{-4}
\int^{t}_{t_{f}}
\left(\frac{a(t^{\prime})}
{a(t_{f})}\right)^{4-3 \gamma}
e^{ -\gamma \Gamma_{\phi} (t^{\prime}-t_f)}
~\gamma \Gamma_{\phi} dt^{\prime}~.
\label{eq:rad}
\end{equation}
For $t_{f} \ll t_{d}$ and $\nu =2$, this expression is
approximated by
\begin{equation}
\rho_{r}(t)=\rho_{f}\left(\frac {t}{t_f}\right)^{-8/3}
\int^{t}_{0}
\left(\frac{t^{\prime}}{t_{f}}\right)^{2/3}
e^{-\Gamma_{\phi}t^{\prime}} dt^{\prime}~,
\label{eq:appr}
\end{equation}
which, using the formula
\begin{equation}
\int_{0}^{u} x^{p-1} e^{-x}dx =
e^{-u}~\sum^{\infty}_{k=0}~
\frac {u^{p+k}}{p(p+1)\cdot\cdot\cdot(p+k)}~~,
\label{eq:formula}
\end{equation}
can be written as
\begin{equation}
\rho_{r} = \frac {3}{5}~\rho~\Gamma_{\phi}t
\left[1 + \frac {3}{8}~\Gamma_{\phi}t
+\frac {9}{88}~(\Gamma_{\phi}t)^2+\cdots\right],
\label{eq:expand}
\end{equation}
with $\rho=\rho_{f} (t/t_{f})^{-2}{\rm{exp}}
(-\Gamma_{\phi}t)$ being the energy density of the field
$\phi$ which performs damped oscillations and decays into
`light' particles.

\par
The energy density of the `new radiation' grows relative
to the energy density of the oscillating field and becomes
essentially equal to it at a cosmic time $t_{d} =
\Gamma_{\phi}^{-1}$ as one can deduce from
(\ref{eq:expand}). After this time, the universe enters
into the radiation dominated era and the normal big bang
\index{big bang}
cosmology is recovered. The temperature at $t_{d}$,
$T_{r}(t_{d})$, is historically called the `reheat'
temperature \index{`reheat' temperature} although no
supercooling and subsequent
reheating \index{reheating} of the universe actually
takes place. Using (\ref{eq:temptime}), we find that
\begin{equation}
T_{r}=\left(\frac {45}{16 \pi^{3}g_*}\right)^{1/4}
(\Gamma_{\phi} M_{P})^{1/2},
\label{eq:reheat}
\end{equation}
where $g_*$ is the effective number of degrees of freedom.
For $V(\phi)=\lambda\phi^{\nu}$,
the total expansion of the universe during the
damped field oscillations is
\begin{equation}
\frac{a(t_{d})}{a(t_{f})}=\left(
\frac{t_{d}}{t_{f}}\right)^{\frac{\nu+2}{3\nu}}.
\label{eq:expansion}
\end{equation}

\section{Density Perturbations
\index{density perturbations} from Inflation
\index{inflation}}
\label{sec:density}

We will now sketch how inflation \index{inflation} solves
the density perturbation \index{density perturbations}
problem described in Sect.\ref{subsec:fluct}. As a matter
of fact, inflation \index{inflation} not only homogenizes
the universe but also provides us with the primordial
density perturbations \index{density perturbations}
needed for structure formation. \index{structure formation}
To understand the origin of these fluctuations, we will
introduce the notion of event horizon. \index{horizon} Our
event horizon, at a cosmic time $t$, includes all points with
which we will eventually communicate sending signals at $t$.
The `instantaneous' (at $t$) radius of the event horizon
\index{horizon} is
\begin{equation}
d_{e}(t)=a(t)\int^{\infty}_{t}
\frac{dt^{\prime}}{a(t^{\prime})}~\cdot
\label{eq:event}
\end{equation}
It is obvious, from this formula, that the event horizon
\index{horizon} is infinite for `matter' or `radiation'. For
inflation, \index{inflation} however, we obtain a slowly
varying event horizon \index{horizon} with $d_{e}(t)=H^{-1}
<\infty$. Points, in our event horizon \index{horizon} at
$t$, with which we can communicate sending signals at $t$,
are eventually pulled away by the exponential expansion and we
cease to be able to communicate with them emitting signals at
later times. We say that these points (and the corresponding
scales) crossed outside the event horizon. \index{horizon}
The situation is similar to that of a black hole. Indeed, the
exponentially expanding (de Sitter) space is like a black hole
turned inside out. We are inside and the black hole surrounds
us from all sides. Then, exactly as in a black hole, there are
quantum fluctuations of the `thermal type' governed by the
Hawking temperature \index{Hawking temperature}
\cite{hawking,gibbons} $T_{H}=H/2\pi$. It turns out
\cite{bunch,vilenkin} that the quantum fluctuations of all
massless fields (the inflaton \index{inflaton} is nearly
massless due to the flatness of the potential) are
$\delta\phi=T_{H}$. These fluctuations of $\phi$ lead to
energy density perturbations \index{density perturbations}
$\delta\rho=V^{\prime}(\phi)\delta\phi$. As the scale
of this perturbations crosses outside the event horizon,
\index{horizon} they become \cite{fischler} classical
metric perturbations.

\par
The evolution of these fluctuations outside the event
horizon \index{horizon} is quite subtle due to the gauge
freedom in general relativity. However, there is a simple
gauge invariant quantity $\zeta\approx\delta\rho/(\rho+p)$
\cite{zeta}, which remains constant outside the horizon.
\index{horizon} Thus, the density perturbation
\index{density perturbations} at any present physical
(`comoving') scale $\ell$, $(\delta\rho/\rho)_{\ell}$,
when this scale crosses inside the post-inflationary particle
horizon \index{horizon} ($p$=0 at this instance) can be
related to the value of $\zeta$ when the same scale crossed
outside the inflationary event horizon \index{horizon} (at
$\ell\sim H^{-1}$). This latter value of $\zeta$ is found,
using (\ref{eq:infeq}), to be
\begin{equation}
\zeta \mid_{\ell\sim H^{-1}}=\left(\frac
{\delta\rho}{\dot{\phi}^2}\right)_{\ell\sim H^{-1}}
=\left(\frac {V^{\prime}(\phi) H(\phi)}
{2 \pi \dot{\phi}^2}\right)_{\ell\sim H^{-1}}
=-\left(\frac {9 H^{3}(\phi)}
{2\pi V^{\prime}(\phi)}\right)_{\ell\sim H^{-1}}\cdot
\label{eq:zeta}
\end{equation}
Taking into account an extra 2/5 factor from the fact that
the universe is matter dominated when the scale $\ell$
re-enters the horizon, \index{horizon} we obtain
\begin{equation}
\left(\frac {\delta\rho}{\rho}\right)_{\ell}=
\frac {16\sqrt{6 \pi}}{5}~\frac
{V^{3/2}(\phi_{\ell})}{M^{3}_{P} V^{\prime}
(\phi_{\ell})}~\cdot
\label{eq:deltarho}
\end{equation}

\par
The calculation of $\phi_{\ell}$, the value of the
inflaton \index{inflaton} field when the `comoving'
scale $\ell$ crossed outside the event horizon,
\index{horizon} goes as follows. A `comoving'
(present physical) scale $\ell$, at $T_r$, was equal to
$\ell(a(t_{d})/a(t_{0}))=\ell(T_{0}/T_{r})$.
Its magnitude at the end of inflation \index{inflation}
($t=t_{f}$) was
equal to $\ell(T_{0}/T_{r})(a(t_{f})/a(t_{d}))$
$=\ell(T_{0}/T_{r})(t_{f}/t_{d})^{(\nu+2)/3 \nu}$
$\equiv\ell_{{\rm{phys}}}(t_{f})$, where the potential
$V(\phi)=\lambda\phi^{\nu}$ was assumed. The scale
$\ell$, when it crossed outside the inflationary horizon,
\index{horizon} was equal to $H^{-1}(\phi_{\ell})$. We,
thus, obtain
\begin{equation}
H^{-1}(\phi_{\ell}) e^{N(\phi_{\ell})} =
\ell_{{\rm{phys}}}(t_{f})~.
\label{eq:lphys}
\end{equation}
Solving this equation, one can calculate $\phi_{\ell}$
and, thus, $N(\phi_{\ell})\equiv N_{\ell}$, the number
of e-foldings the scale $\ell$ suffered during inflation.
\index{inflation} In particular, for our present horizon
\index{horizon} scale $\ell\approx 2H_{0}^{-1}\sim
10^4~{\rm Mpc}$, it turns out that $N_{H_{0}}\approx
50-60$.

\par
Taking the potential $V(\phi)=\lambda \phi^4$,
(\ref{eq:expefold}), (\ref{eq:deltarho}) and
(\ref{eq:lphys}) give
\begin{equation}
\left(\frac{\delta \rho} {\rho}\right)_{\ell}=
\frac {4 \sqrt{6 \pi}} {5}\lambda^{1/2}
\left(\frac{\phi_{\ell}}{M_{P}}\right)^3 =
\frac {4 \sqrt{6 \pi}}{5}\lambda^{1/2}
\left(\frac {N_{\ell}}{\pi}\right)^{3/2}\cdot
\label{eq:nl}
\end{equation}
From the result of COBE \index{COBE} \cite{cobe},
$(\delta\rho/\rho)_{H_{0}}\approx 6\times
10^{-5}$, one can then deduce that $\lambda\approx 6\times
10^{-14}$ for $N_{H_{0}}\approx 55$. We thus see that the
inflaton \index{inflaton} must be a very weakly coupled field.
In non-SUSY \index{SUSY} GUTs, \index{GUT} the inflaton
\index{inflaton} is necessarily gauge singlet since otherwise
radiative corrections will make it strongly coupled. This is
not so satisfactory since it forces us to introduce an
otherwise unmotivated very weakly coupled gauge singlet. In
SUSY \index{SUSY} GUTs, \index{GUT} however, the inflaton
\index{inflaton} could be identified \cite{nonsinglet}
with a conjugate pair of gauge non-singlet fields $\phi$,
$\bar{\phi}$ already present in the theory and causing the
gauge symmetry breaking. Absence of strong radiative
corrections from gauge interactions is guaranteed by the
mutual cancellation of the D-terms of these fields.

\par
The spectrum of density perturbations
\index{density perturbations}which emerge from inflation
\index{inflation} can also be analyzed. We will again take
the potential $V(\phi)=\lambda \phi^{\nu}$. One then
finds that $(\delta \rho/ \rho)_{\ell}$ is proportional
to $\phi_{\ell}^{(\nu+2)/2}$ which, combined with the
fact that $N(\phi_{\ell})$ is proportional to
$\phi_{\ell}^{2}$ (see (\ref{eq:expefold})), gives
\begin{equation}
\left(\frac {\delta \rho}{\rho}\right)_{\ell} =
\left(\frac{\delta \rho}{\rho}\right)_{H_{0}}\left(
\frac{N_{\ell}}{N_{H_{0}}}\right)^{\frac{\nu+2}{4}}.
\label{eq:spectrum}
\end{equation}
The scale $\ell$ divided by the size of our present horizon
\index{horizon} ($\approx 10^4~{\rm{Mpc}}$) should equal
${\rm exp}(N_{\ell}-N_{H_{0}})$. This gives $N_{\ell}/
N_{H_{0}}=1+\ln(\ell/10^{4})^{1/N_{H_{0}}}$ which expanded
around $\ell\approx 10^4~{\rm Mpc}$ and substituted in
(\ref{eq:spectrum}) yields
\begin{equation}
\left(\frac{\delta \rho}{\rho}\right)_{\ell}\approx
\left(\frac{\delta \rho}{\rho}\right)_{H_{0}}\left(
\frac {\ell}{10^4~{\rm{Mpc}}}\right)^{\alpha_{s}},
\label{eq:alphas}
\end{equation}
with $\alpha_{s}=(\nu+2)/4N_{H_{0}}$. For $\nu=4$,
$\alpha_{s}\approx 0.03$ and, thus,  the density
perturbations \index{density perturbations} are
essentially scale independent.

\section{Density Perturbations
\index{density perturbations} in `Matter'}
\label{sec:matter}

We will now discuss the evolution of the primordial density
perturbations \index{density perturbations} after their
scale enters the post-inflationary horizon. \index{horizon}
To this end, we introduce \cite{bardeen} the `conformal
time', $\eta$, so that the Robertson-Walker metric takes the
form of a conformally expanding Minkowski space:
\begin{equation}
ds^2=-dt^2+a^2(t)~d\bar{r}^2=a^{2}(\eta)~
(-d\eta^2+d\bar{r}^2)~,
\label{eq:conf}
\end{equation}
where $\bar{r}$ is a `comoving' 3-vector. The Hubble
parameter now takes the form $H\equiv\dot{a}(t)/a(t)
=a^{\prime}(\eta)/a^{2}(\eta)$ and the Friedmann equation
\index{Friedmann equation} (\ref{eq:friedmann}) is rewritten
as
\begin{equation}
\frac{1}{a^{2}}\left(\frac{a^{\prime}}{a}\right)^{2}
=\frac{8\pi G}{3}\rho~,
\label{eq:conffried}
\end{equation}
where primes denote derivation with respect to $\eta$. The
continuity equation (\ref{eq:continuity}) takes the form
$\rho^{\prime}=-3\tilde{H}(\rho+p)$ with $\tilde{H}=
a^{\prime}/a$. For `matter', $\rho\propto a^{-3}$ which
gives $a=(\eta/\eta_{0})^2$ and $a^{\prime}/a=2/\eta$
($\eta_0$ is the present value of $\eta$).

\par
The Newtonian equation \index{Newtonian equation}
(\ref{eq:newton}) can now be written in
the form
\begin{equation}
\delta^{\prime\prime}_{\bar{k}}(\eta) +
\frac {a^{\prime}}{a}
\delta^{\prime}_{\bar{k}}(\eta)
- 4\pi G\rho a^{2}\delta_{\bar{k}} (\eta) =0~,
\label{eq:confnewton}
\end{equation}
and the growing (Jeans unstable) mode
$\delta_{\bar{k}}(\eta)\propto\eta^{2}$ and is
expressed \cite{schaefer} as
\begin{equation}
\delta_{\bar{k}}(\eta)=\epsilon_{H}\left(
\frac{k\eta}{2}\right)^{2}\hat {s}(\bar{k})~,
\label{eq:growmode}
\end{equation}
where $\hat{s}(\bar{k})$ is a Gaussian random variable
satisfying
\begin{equation}
<\hat{s}(\bar {k})>=0~,~<\hat{s}(\bar {k})
\hat{s}(\bar {k}^{\prime})>
=\frac {1}{k^{3}}\delta(\bar{k}-
\bar{k}^{\prime})~,
\label{eq:gauss}
\end{equation}
and $\epsilon_{H}$ is the amplitude of the perturbation
when its scale crosses inside the post-inflationary horizon.
\index{horizon} The latter can be seen as follows. A
`comoving' (present physical) length $\ell$ crosses inside
the post-inflationary horizon \index{horizon} when
$a\ell/2 \pi=H^{-1}=a^2/a^{\prime}$ which gives
$\ell/2\pi\equiv k^{-1}=a/a^{\prime}=\eta_{H}/2$ or
$k \eta_{H}/2 = 1$, where $\eta_{H}$ is the `conformal
time' at horizon \index{horizon} crossing. This means that,
at horizon crossing, $\delta_{\bar{k}}(\eta_{H})=
\epsilon_{H}\hat{s}(\bar{k})$. For scale invariant
perturbations, the amplitude $\epsilon _{H}$ is constant.
The gauge invariant perturbations of the scalar gravitational
potential are given \cite{bardeen} by the Poisson's equation
\begin{equation}
\Phi=-4\pi G\frac{a^2}{k^2}\rho
\delta_{\bar{k}}(\eta)~.
\label{eq:poisson}
\end{equation}
From the Friedmann equation \index{Friedmann equation}
(\ref{eq:conffried}), we then obtain
\begin{equation}
\Phi=-\frac{3}{2}\epsilon_{H}\hat{s}(\bar{k})~.
\label{eq:scalarpot}
\end{equation}

\par
The spectrum of the density perturbations
\index{density perturbations} can be characterized by the
correlation function ($\bar{x}$ is a `comoving' 3-vector)
\begin{equation}
\xi(\bar{r})\equiv
<\tilde{\delta}^{*}(\bar{x},\eta)
\tilde{\delta}(\bar{x}+\bar{r},
\eta)>~,
\label{eq:corr}
\end{equation}
where
\begin{equation}
\tilde{\delta}(\bar{x},\eta)=\int d^{3}k
\delta_{\bar{k}}(\eta)e^{i\bar{k}\bar{x}}.
\label{eq:fourier}
\end{equation}
Substituting (\ref{eq:growmode}) in (\ref{eq:corr})
and then using (\ref{eq:gauss}), we obtain
\begin{equation}
\xi (\bar{r})=\int d^{3}ke^{-i\bar{k}\bar{r}}
\epsilon^2_{H}
\left(\frac {k \eta}{2}\right)^{4}\frac{1}{k^{3}}~,
\label{eq:index}
\end{equation}
and the spectral function  $P(k,\eta)=
\epsilon^2_{H}(\eta^{4}/16)k$ is proportional to $k$
for $\epsilon_{H}$ constant. We say that, in this case,
the spectral index \index{spectral index} $n=1$ and we
have a Harrison-Zeldovich \cite{hz} flat spectrum. In the
general case, $P\propto k^{n}$ with $n = 1-2\alpha_{s}$
(see (\ref{eq:alphas})). For $V(\phi)=\lambda\phi^{4}$,
we get $n\approx 0.94$.

\section{Temperature Fluctuations
\index{temperature fluctuations}}
\label{sec:temperature}

The density inhomogeneities produce temperature fluctuations
\index{temperature fluctuations} in the CMBR. \index{CMBR}
For angles $\theta\stackrel{_{>}}{_{\sim }}2^{o}$, the
dominant effect is the scalar Sachs-Wolfe \cite{sachswolfe}
effect. \index{Sachs-Wolfe effect} Density perturbations
\index{density perturbations} on the `last scattering
surface' cause scalar gravitational potential fluctuations,
$\Phi$, which then produce temperature fluctuations
\index{temperature fluctuations} in the CMBR. \index{CMBR}
The reason is that regions with a deep gravitational
potential will cause the photons to lose energy as they
climb up the well and, thus, appear cooler. For
$\theta\stackrel{_{<}}{_{\sim }}2^{o}$, the dominant
effects are: i) Motion of the `last scattering surface'
\index{last scattering surface} causing Doppler shifts,
and ii) Intrinsic fluctuations of the photon temperature
which are more difficult to calculate since they depend
on microphysics, the ionization history, photon streaming
and other effects.

\par
The temperature fluctuations \index{temperature fluctuations}
at an angle $\theta$ due to the scalar Sachs-Wolfe effect
\index{Sachs-Wolfe effect} turn out \cite{sachswolfe}
to be  $(\delta T/T)_{\theta}=-\Phi_{\ell}/3$, with
$\ell$ being the `comoving' scale on the `last scattering
surface' which subtends the angle $\theta$
$[~\ell\approx 100~h^{-1}(\theta/{\rm{degrees}})
~{\rm{Mpc}}~]$ and $\Phi_{\ell}$ the corresponding
scalar gravitational potential fluctuations. From
(\ref{eq:scalarpot}), we then obtain
$(\delta T/T)_{\theta}=
(\epsilon_{H}/2)\hat{s}(\bar{k})$, which using
(\ref{eq:growmode}) gives the relation
\begin{equation}
\left(\frac{\delta T}{T}\right)_{\theta}=
\frac{1}{2} \delta_{\bar{k}}(\eta_{H})
=\frac {1}{2}\left(\frac{\delta \rho}
{\rho}\right)_{\ell\sim 2\pi k^{-1}}\cdot
\label{eq:swe}
\end{equation}
The COBE \index{COBE} scale (present horizon) \index{horizon}
corresponds to $\theta\approx 60^{o}$. Equations
(\ref{eq:expefold}), (\ref{eq:deltarho}) and (\ref{eq:swe})
give
\begin{equation}
\left(\frac {\delta T}{T}\right)_{\ell}
\propto\left(\frac{\delta\rho}{\rho}\right)_{\ell}
\propto
\frac{V^{3/2}(\phi_{\ell})}{M^{3}_{P} V^{\prime}
(\phi_{\ell})}\propto N_{\ell}^{\frac{\nu+2}{4}}.
\label{ewq:tempflu}
\end{equation}
Analyzing the temperature fluctuations
\index{temperature fluctuations} in spherical harmonics,
one can obtain the quadrupole anisotropy
\index{quadrupole anisotropy} due to the scalar
Sachs-Wolfe effect: \index{Sachs-Wolfe effect}
\begin{equation}
\left(\frac{\delta T}{T}\right)_{Q-S} =
\left(\frac{32 \pi}
{45}\right)^{1/2}
\frac{V^{3/2}(\phi_{\ell})}{M^{3}_{P}
V^{\prime}(\phi_{\ell})}~\cdot
\label{eq:quadrupole}
\end{equation}
For $V(\phi)=\lambda\phi^{\nu}$, this becomes
\begin{equation}
\left(\frac{\delta T}{T}\right)_{Q-S} =
\left(\frac{32 \pi}{45}\right)^{1/2}
\frac{\lambda^{1/2}\phi_{\ell}^{\frac{\nu+2}{2}}}
{\nu M^{3}_{P}}=
\left(\frac{32 \pi}{45}\right)^{1/2}
\frac {\lambda^{1/2}}{\nu M^{3}_{P}}
\left(\frac{\nu M^{2}_{P}}
{4 \pi}\right)^{\frac{\nu+2}{4}}
N_{\ell}^{\frac{\nu+2}{4}}.
\label{eq:anisotropy}
\end{equation}
Comparing this with the COBE \index{COBE} \cite{cobe}
result, $(\delta T/T)_{Q}\approx 6.6\times 10^{-6}$,
we obtain $\lambda\approx 6\times 10^{-14}$ for
$\nu=4$ and number of e-foldings suffered by our
present horizon \index{horizon} scale during the
inflationary phase  $N_{\ell\sim H^{-1}_{0}}\equiv N_{Q}
\approx 55$.

\par
There are also `tensor' fluctuations \cite{tensor}
\index{`tensor' fluctuations} in the temperature of the CMBR.
\index{CMBR} The `tensor' quadrupole anisotropy
\index{quadrupole anisotropy} is
\begin{equation}
\left(\frac{\delta T}{T}\right)_{Q-T}
\approx 0.77~\frac
{V^{1/2}(\phi_{\ell})}{M^{2}_{P}}~\cdot
\label{eq:tensor}
\end{equation}
The total quadrupole anisotropy
\index{quadrupole anisotropy} is given by
\begin{equation}
\left(\frac{\delta T}{T}\right)_{Q}=
\left[\left(\frac{\delta T}{T}\right)^{2}_{Q-S}+
\left(\frac{\delta T}{T}\right)^{2}_{Q-T}
\right]^{1/2},
\label{eq:total}
\end{equation}
and the ratio
\begin{equation}
r=\frac{\left(\delta T/T\right)^{2}_{Q-T}}
{\left(\delta T/T\right)^{2}_{Q-S}}\approx 0.27
~\left(\frac{M_{P}V^{\prime}(\phi_{\ell})}
{V(\phi_{\ell})}\right)^{2}\cdot
\label{eq:ratio}
\end{equation}
For $V(\phi)=\lambda\phi^{\nu}$, we obtain
$r\approx 3.4~\nu/N_{H}\ll 1$, and the `tensor'
contribution to the temperature fluctuations
\index{temperature fluctuations} of the CMBR
\index{CMBR} is negligible.

\section{Hybrid Inflation \index{hybrid inflation}}
\label{sec:hybrid}

\subsection{The non-Supersymmetric Version}
\label{subsec:nonsusy}

The basic disadvantage of inflationary scenarios such as the
`new' \cite{new} or `chaotic' \cite{chaotic} ones is that
they require tiny coupling constants in order to reproduce
the results of COBE \index{COBE} \cite{cobe}. This has led
Linde \cite{hybrid} to propose, in the context of non-SUSY
\index{SUSY} GUTs, \index{GUT} the hybrid inflationary
scenario. The idea was to use two real scalar fields $\chi$
and $\sigma$ instead of one that was normally used. $\chi$
provides the `vacuum' energy density which drives inflation,
\index{inflation} while $\sigma$ is the slowly varying
field during inflation. \index{inflation} This splitting of
roles between two fields allows us to reproduce the COBE
\index{COBE} results with `natural' (not too small) values
of the relevant parameters in contrast to previous realizations
of inflation. \index{inflation}

\par
The scalar potential utilized by Linde is
\begin{equation}
V(\chi,\sigma)=\kappa^2 \left(M^2-
\frac{\chi^2}{4}\right)^2+
\frac{\lambda^2\chi^2 \sigma^2}{4}
+\frac {m^2\sigma^2}{2}~,
\label{eq:lindepot}
\end{equation}
where $\kappa$, $\lambda$ are dimensionless positive
coupling constants and $M$, $m$ are mass parameters.
The vacua lie at $\langle\chi\rangle=\pm 2M$,
$\langle\sigma\rangle=0$. Putting $m$=0, we see
that $V$ possesses a flat direction at $\chi=0$ with
$V(\chi=0,\sigma)=\kappa^2M^4$. The ${\rm mass}^2$
of $\chi$ along this direction is
$m^2_\chi=-\kappa^2M^2+\lambda^2\sigma^2/2$. So,
for $\chi=0$ and $\vert\sigma\vert>\sigma_c=
\sqrt{2}\kappa M/\lambda$, we obtain a flat valley
of minima. Reintroducing $m\neq 0$, this valley acquires
a non-zero slope and the system can inflate as it rolls
down this valley. This scenario is called hybrid since
the `vacuum' energy density ($\approx\kappa^2M^4$) is
provided by $\chi$, while the slowly rolling field is
$\sigma$.

\par
The $\epsilon$ and $\eta$ criteria (see (\ref{eq:src}))
imply that, for the relevant values of parameters (see
below), inflation \index{inflation} continues until
$\sigma$ reaches $\sigma_c$, where it terminates abruptly.
It is followed by a `waterfall', i.e., a sudden entrance
into an oscillatory phase about a global minimum. Since
the system can fall into either of the two minima with
equal probability, topological defects (monopoles,
\index{monopole} cosmic strings \index{cosmic strings}
or domain walls) \index{domain walls} are copiously
produced \cite{smooth} if they are predicted by the
particular particle physics model employed. So, if the
underlying GUT \index{GUT} gauge symmetry breaking (by
$\langle\chi\rangle$) leads to the existence of monopoles
\index{monopole} or domain walls, \index{domain walls} we
encounter a cosmological catastrophe.

\par
The onset of hybrid inflation \index{hybrid inflation}
requires \cite{onset} that, at $t\sim H^{-1}$, $H$ being
the inflationary Hubble parameter, a region exists with
size $\stackrel{_{>}}{_{\sim}}H^{-1}$, where $\chi$
and $\sigma$ are almost uniform with negligible kinetic
energies and values close to the bottom of the valley of
minima. Such a region, at $t_P$, would have been much
larger than the Planck length $\ell_P$ and it is, thus,
difficult to imagine how it could be so homogeneous.
Moreover, as it has been argued \cite{initial}, the
initial values (at $t_P$) of the fields in this region
must be strongly restricted in order to obtain adequate
inflation. \index{inflation} Several possible solutions
to this problem of initial conditions for hybrid inflation
\index{hybrid inflation} have been proposed (see e.g.,
\cite{double,sugra,costas}).

\par
The quadrupole anisotropy \index{quadrupole anisotropy}
of the CMBR \index{CMBR} produced during hybrid inflation
\index{hybrid inflation} can be estimated, using
(\ref{eq:quadrupole}), to be
\begin{equation}
\left(\frac{\delta T}{T}\right)_{Q}\approx
\left(\frac{16\pi}{45}\right)^{\frac{1}{2}}
\frac{\lambda\kappa^2M^5}{M^3_Pm^2}~\cdot
\label{eq:lindetemp}
\end{equation}
The COBE \index{COBE} \cite{cobe} result,
$(\delta T/T)_{Q}\approx 6.6\times 10^{-6}$, can then be
reproduced with $M\approx 2.86\times 10^{16}~{\rm GeV}$,
the SUSY \index{SUSY} GUT \index{GUT} vev, and
$m\approx 1.3~\kappa\sqrt{\lambda}
\times 10^{15}~{\rm GeV}$. Note that
$m\sim 10^{12}~{\rm GeV}$ for $\kappa$,
$\lambda\sim 10^{-2}$.

\subsection{The Supersymmetric Version}
\label{subsec:susy}

Hybrid inflation \index{hybrid inflation} is \cite{lyth}
`tailor made' for globally SUSY \index{SUSY} GUTs \index{GUT}
except that an intermediate scale mass for $\sigma$ cannot be
obtained. Actually, all scalars acquire masses
$\sim m_{3/2} \sim 1~{\rm TeV}$ (the gravitino mass)
from soft SUSY \index{SUSY} breaking.

\par
Let us consider the renormalizable superpotential
\begin{equation}
W=\kappa S(-M^2+\bar{\phi}\phi)~,
\label{eq:superpot}
\end {equation}
where $\bar{\phi}$, $\phi$ is a  pair of $G_{S}$ singlet
left handed superfields belonging to non-trivial conjugate
representations of the GUT \index{GUT} gauge group $G$ and
reducing its rank by their vevs, and $S$ is a gauge singlet
left handed superfield. The parameters $\kappa$ and $M$
($\sim 10^{16}~{\rm GeV}$) are made positive by
field redefinitions. The vanishing of the F-term $F_S$ gives
$\langle\bar{\phi}\rangle\langle\phi\rangle=M^2$,
and the D-terms vanish for $\vert\langle\bar{\phi}
\rangle\vert=\vert\langle\phi\rangle\vert$. So, the
SUSY \index{SUSY} vacua lie at $\langle\bar{\phi}
\rangle^*=\langle\phi\rangle=\pm M$ and $\langle S
\rangle=0$ (from $F_{\bar{\phi}}=F_{\phi}=0$). We see
that $W$ leads to the spontaneous breaking of $G$.

\par
$W$ also gives rise to hybrid inflation.
\index{hybrid inflation} The potential derived from it is
\begin{equation}
V(\bar{\phi},\phi,S)=
\kappa^2\vert M^2-\bar{\phi}\phi\vert^2+
\kappa^2\vert S\vert^2(\vert\bar{\phi}\vert^2+
\vert\phi\vert^2)+{\rm{D-terms}}~.
\label{eq:hybpot}
\end{equation}
D-flatness implies $\bar{\phi}^*=e^{i\theta}\phi$. We
take $\theta=0$, so that the SUSY \index{SUSY} vacua are
contained. $W$ has a $U(1)_R$ R-symmetry:
$\bar{\phi}\phi\to\bar{\phi}\phi$, $S\to e^{i\alpha}S$,
$W\to e^{i\alpha}W$. Actually, $W$ is the most general
renormalizable superpotential allowed by $G$ and $U(1)_R$.
Bringing $\bar{\phi}$, $\phi$, $S$ on the real axis by $G$
and $U(1)_R$ transformations, we write
$\bar{\phi}=\phi\equiv\chi/2$, $S\equiv\sigma/\sqrt{2}$
where $\chi$, $\sigma$ are normalized real scalar fields.
$V$ then takes the form in (\ref{eq:lindepot}) with
$\kappa=\lambda$ and $m=0$. So, Linde's potential is almost
obtainable from SUSY \index{SUSY} GUTs \index{GUT} but without
the mass term of $\sigma$ which is, however, crucial for driving
the inflaton \index{inflaton} towards the vacua.

\par
One way to generate a slope along the inflationary valley
($\bar{\phi}=\phi=0$, $\vert S\vert>S_{c}\equiv M$) is
\cite{dss} to include the one-loop radiative corrections. In
fact, SUSY \index{SUSY} breaking by the `vacuum' energy density
$\kappa^2M^4$ along this valley causes a mass splitting in
the supermultiplets $\bar{\phi}$, $\phi$. We obtain a
Dirac fermion with ${\rm mass}^2=\kappa^2\vert S\vert^2$
and two complex scalars with ${\rm mass}^2=
\kappa^2\vert S\vert^2\pm\kappa^2M^2$. This leads to
the existence of one-loop radiative corrections to $V$
on the inflationary valley which are found
from the Coleman-Weinberg formula \cite{cw}:
\begin{equation}
\Delta V=\frac{1}{64\pi^2}\sum_i(-)^{F_i}\ M_i^4\ln
\frac{M_i^2}{\Lambda^2}~,
\label{eq:deltav}
\end{equation}
where the sum extends over all helicity states $i$, with
fermion number $F_i$ and ${\rm mass}^2=M_i^2$, and
$\Lambda$ is a renormalization scale. We find that
$\Delta V(\vert S\vert)$ is
\begin{equation}
\kappa^2 M^4~{\kappa^2N\over 32\pi^2}\left(
2\ln{\kappa^2\vert S\vert^2\over\Lambda^2}
+(z+1)^{2}\ln(1+z^{-1})+(z-1)^{2}\ln(1-z^{-1})\right),
\label{eq:rc}
\end{equation}
where $z=x^2=\vert S\vert^2/M^2$ and $N$ is the
dimensionality of the representations to which $\bar{\phi}$,
$\phi$ belong. For $z\gg 1$ ($\vert S\vert\gg S_c$), the
effective potential on the inflationary valley can be expanded
as
\cite{dss,lss}
\begin{equation}
V_{{\rm{eff}}}(\vert S\vert)=\kappa^2 M^4
\left[1+\frac{\kappa^2N}{16\pi^2}\left(\ln
\frac{\kappa^2\vert S\vert^2}{\Lambda^2}
+\frac{3}{2}-\frac{1}{12z^2}+\cdots\right)\right].
\label{eq:veff}
\end{equation}
The slope on this valley from these radiative corrections
is $\Lambda$-independent.

\par
From (\ref{eq:expefold}), (\ref{eq:quadrupole}) and
(\ref{eq:rc}), we find the quadrupole anisotropy
\index{quadrupole anisotropy} of the CMBR: \index{CMBR}
\begin{equation}
\left(\frac{\delta T}{T}\right)_{Q}\approx\frac{8\pi}
{\sqrt{N}}\left(\frac{N_{Q}}{45}\right)^{\frac{1}{2}}
\left(\frac{M}{M_{P}}\right)^2x_Q^{-1}y_Q^{-1}
\Lambda(x_Q^2)^{-1},
\label{eq:qa}
\end{equation}
with
\begin{equation}
\Lambda(z)=(z+1)\ln(1+z^{-1})+(z-1)\ln(1-z^{-1})~,
\label{eq:lambda}
\end{equation}
\begin{equation}
y_Q^2=\int_1^{x_Q^2}\frac{dz}{z}\Lambda(z)^{-1},
~y_Q\geq 0~.
\label{eq:yq}
\end{equation}
Here, $N_{Q}$ is the number of e-foldings suffered by our
present horizon \index{horizon} scale during inflation,
\index{inflation} and $x_Q=\vert S_Q\vert/M$, with $S_Q$
being the value of $S$ when our present horizon
\index{horizon} scale crossed outside the inflationary
horizon. \index{horizon} For $\vert S_Q\vert\gg S_c$, $y_Q=
x_Q(1-7/12x_Q^2+\cdots)$. Finally, from
(\ref{eq:rc}), one finds
\begin{equation}
\kappa\approx\frac{8\pi^{\frac{3}{2}}}{\sqrt{NN_Q}}
~y_Q~\frac{M}{M_{P}}~\cdot
\label{eq:kappa}
\end{equation}

\par
The slow roll conditions (see (\ref{eq:src})) for SUSY
\index{SUSY} hybrid inflation \index{hybrid inflation} are
$\epsilon,\vert\eta\vert\leq 1$, where
\begin{equation}
\epsilon=\left(\frac{\kappa^2M_P}{16\pi^2M}\right)^2
\frac{N^2x^2}{8\pi}\Lambda(x^2)^2,
\label{eq:epsilon}
\end{equation}
\begin{equation}
\eta=\left(\frac{\kappa M_P}{4\pi M}\right)^2
\frac{N}{8\pi}\left((3z+1)\ln(1+z^{-1})+
(3z-1)\ln(1-z^{-1})\right).
\label{eq:eta}
\end{equation}
Note that $\eta\rightarrow-\infty$ as $x\rightarrow 1^+$.
However, for most relevant values of the parameters
($\kappa\ll 1)$, the slow roll conditions are violated only
`infinitesimally' close to the critical point at $x=1$
($\vert S\vert=S_c$). So, inflation \index{inflation}
continues practically
until this point is reaches, where the `waterfall' occurs.

\par
From the COBE \index{COBE} \cite{cobe} result,
$(\delta T/T)_{Q}\approx 6.6\times 10^{-6}$, and
eliminating $x_Q$ between (\ref{eq:qa}) and
(\ref{eq:kappa}), we obtain $M$ as a function of $\kappa$.
For $x_Q\rightarrow\infty$, $y_Q\rightarrow x_Q$
and $x_Qy_Q\Lambda(x_Q^2)\rightarrow 1^-$. Thus, the
maximal $M$ is achieved in this limit and equals about
$10^{16}~{\rm GeV}$ (for $N=8$, $N_Q\approx 55)$. This
value of $M$, although somewhat smaller than the SUSY
\index{SUSY} GUT \index{GUT} scale, is quite close to it. As
a numerical example, take $\kappa=4\times 10^{-3}$ which gives
$M\approx 9.57\times 10^{15}~{\rm GeV}$, $x_Q\approx2.633$,
$y_Q\approx 2.42$. The slow roll conditions are violated at
$x-1\approx 7.23\times 10^{-5}$, where $\eta=-1$ ($\epsilon
\approx 8.17\times 10^{-8}$ at $x=1$). The spectral index
\index{spectral index} of density perturbations
\index{density perturbations} $n=1-6\epsilon+2\eta$
\cite{liddle} is about 0.985.

\par
SUSY \index{SUSY} hybrid inflation \index{hybrid inflation}
is considered `natural' for the following reasons:
\begin{list}
\setlength{\rightmargin=0cm}{\leftmargin=0cm}
\item[{\bf i.}] There is no need of tiny coupling constants
($\kappa\sim 10^{-3}$).
\vspace{.25cm}
\item[{\bf ii.}] $W$ in (\ref{eq:superpot}) has the most
general renormalizable form allowed by $G$ and $U(1)_R$. The
coexistence of the $S$ and $S\bar{\phi}\phi$ terms in $W$
implies that the combination $\bar{\phi}\phi$ is `neutral'
under all symmetries of $W$ and, thus, all the
non-renormalizable terms of the form $S(\bar{\phi}\phi)^n$,
$n\geq 2$, are also allowed \cite{jean}. The leading term of
this type $S(\bar{\phi}\phi)^2$, if its dimensionless
coefficient is of order unity, can be comparable to
$S\bar{\phi}\phi$ (recall that $\kappa\sim 10^{-3}$) and,
thus, play a role in inflation \index{inflation}
(see Sect.\ref{sec:extensions}). All higher order terms of
this type with $n\geq 3$ give negligible contributions to the
inflationary potential. Note that $U(1)_R$ guarantees the
linearity of $W$ in $S$ to all orders excluding terms such as
$S^2$ which could generate an inflaton \index{inflaton} mass
$\stackrel{_{>}}{_{\sim }}H$ and ruin inflation
\index{inflation} by violating the slow roll conditions.
\vspace{.25cm}
\item[{\bf iii.}] SUSY \index{SUSY} guarantees that the
radiative corrections do not invalidate \cite{nonsinglet}
inflation, \index{inflation} but rather provide \cite{dss}
a slope along the inflationary trajectory, needed for driving
the inflaton \index{inflaton} towards the SUSY \index{SUSY}
vacua.
\vspace{.25cm}
\item[{\bf iv.}] Supergravity (SUGRA) corrections can be
brought under control leaving inflation \index{inflation}
intact. The scalar potential in SUGRA is given \cite{superg}
by
\begin{equation}
V=\exp\left(\frac{K}{m_P^2}\right)\left[\left(
K^{-1}\right)_i^{~j}F^i F_j-3\frac{\left|W\right|^2}
{m_P^2}\right],
\label{eq:sugra}
\end{equation}
where $K$ is the K\"ahler potential, $m_P=M_P/\sqrt{8\pi}
\approx 2.44\times 10^{18}~{\rm GeV}$ is the `reduced'
Planck scale, $F^i=W^i+K^iW/m_P^2$, and upper (lower)
indices denote derivation with respect to the scalar
field $\phi_i$ ($\phi^{j*}$). $K$ is expanded as $K=\vert
S\vert^2+\vert\bar{\phi}\vert^2+\vert\phi\vert^2+\alpha
\vert S\vert^4/m_P^2+\cdots$, where the quadratic terms
constitute the `minimal' K\"ahler potential. The term $\vert
S\vert^2$, whose coefficient is normalized to unity, could
generate a ${\rm mass}^2\sim\kappa^2M^4/m_P^2\sim H^2$ for
$S$ on the inflationary path from the expansion of the
exponential prefactor in (\ref{eq:sugra}). This would ruin
inflation. \index{inflation} Fortunately, with this form of
$W$ (including all the higher order terms), this
${\rm mass}^2$ is cancelled in $V$ \cite{lyth,stewart}. The
linearity of $W$ in $S$, guaranteed to all orders by $U(1)_R$,
is crucial for this cancellation. The $\vert S\vert^4$ term
in $K$ also generates a ${\rm mass}^2$ for $S$ via the factor
$(\partial^2 K/\partial S\partial S^*)^{-1}=1-4\alpha
\vert S\vert^2/m_P^2+\cdots$ in (\ref{eq:sugra}),
which is however not cancelled (see e.g., \cite{quasi}).
In order to avoid ruining inflation, \index{inflation} one
has then to assume \cite{sugra,lss} that $|\alpha|
\stackrel{_{<}}{_{\sim }}10^{-3}$. All other higher
order terms in $K$ give suppressed contributions on the
inflationary path (since $\vert S\vert\ll m_P$). So, we
see that a mild tuning of just one parameter is adequate for
controlling SUGRA corrections. (In other models, tuning of
infinitely many parameters is required.) Moreover, note that
with special forms of $K$ one can solve this problem even
without a mild tuning. An example is given in \cite{costas},
where the dangerous ${\rm{mass}}^2$ could be cancelled in
the presence of fields without superpotential but with large
vevs generated via D-terms. These properties practically
persist even in the extensions of the model we will consider
in Sect.\ref{sec:extensions}.
\end{list}
In summary, for all these reasons, we consider SUSY
\index{SUSY} hybrid inflation \index{hybrid inflation}
(with its extensions) as an extremely `natural' inflationary
scenario.

\section{Extensions of Supersymmetric Hybrid Inflation
\index{hybrid inflation}}
\label{sec:extensions}

In trying to apply (SUSY) \index{SUSY} hybrid inflation
\index{hybrid inflation} to higher GUT \index{GUT} gauge
groups which predict the existence of monopoles,
\index{monopole} we encounter the following problem.
Inflation \index{inflation} is terminated abruptly as the
system reaches the critical point on the inflationary path
and is followed by the `waterfall' regime during which the
scalar fields $\bar\phi$, $\phi$ develop their vevs
starting from zero and the spontaneous breaking of the
GUT \index{GUT}  gauge symmetry takes place. The fields
$\bar\phi$, $\phi$ can end up at any point of the vacuum
manifold with equal probability and, thus, monopoles
\index{monopole} are copiously produced \cite{smooth} via
the Kibble mechanism \index{Kibble mechanism} \cite{kibble}
leading to a cosmological disaster (see e.g., \cite{mono}).

\par
One of the simplest GUTs \index{GUT} predicting monopoles
\index{monopole} is the Pati-Salam (PS) model \index{PS model}
\cite{ps} with gauge group $G_{PS}=SU(4)_c\times SU(2)_L
\times SU(2)_R$. These monopoles \index{monopole} carry two
units of `Dirac' magnetic charge \cite{magg}. We will
present solutions \cite{smooth,jean} of the monopole
\index{monopole} problem of hybrid inflation
\index{hybrid inflation} within the SUSY \index{SUSY} PS
model, \index{PS model} although our mechanisms can be extended
to other semi-simple gauge groups such as the `trinification'
group $SU(3)_c\times SU(3)_L\times SU(3)_R$, which
emerges from string theory and predicts
\cite{trinification,tomaras} monopoles \index{monopole}
with triple `Dirac' charge, and possibly to simple
gauge groups such as $SO(10)$.

\subsection{Shifted Hybrid Inflation
\index{shifted hybrid inflation}}
\label{subsec:shifted}

One idea \cite{jean} for solving the magnetic monopole
\index{monopole} problem is to include into the standard
superpotential for hybrid inflation \index{hybrid inflation}
(shown in (\ref{eq:superpot})) the leading
non-renormalizable term, which, as explained in
Sect.\ref{subsec:susy}, cannot be excluded by any symmetries.
If its dimensionless coefficient is of order unity, this
term can compete with the trilinear coupling of the standard
superpotential (whose coefficient is $\sim 10^{-3}$). The
coexistence of these terms reveals a completely new picture.
In particular, there appears a non-trivial (classically)
flat direction along which $G_{PS}$ is spontaneously broken
with the appropriate Higgs fields acquiring constant values.
This `shifted' flat direction can be used as inflationary
trajectory with the necessary slope obtained again from
one-loop radiative corrections \cite{dss}. The termination
of inflation \index{inflation} is again abrupt followed by
a `waterfall' but no monopoles \index{monopole} are formed
in this transition since $G_{PS}$ is already spontaneously
broken during inflation. \index{inflation}

\par
The spontaneous breaking of the gauge group $G_{PS}$ to $G_S$
is achieved via the vevs of a conjugate pair of Higgs superfields
\begin{eqnarray}
\bar{H}^c &=& (4,1,2) \equiv \left(\begin{array}{cccc}
                       \bar{u}^c_H & \bar{u}^c_H &
                       \bar{u}^c_H & \bar{\nu}_H^c\\
                      \bar{d}^c_H & \bar{d}^c_H &
                      \bar{d}^c_H & \bar{e}^c_H
                      \end{array}\right),\nonumber\\
H^c &=& (\bar{4},1,2) \equiv \left(\begin{array}{cccc}
                       u^c_H & u^c_H & u^c_H & \nu_H^c\\
                       d^c_H & d^c_H & d^c_H & e^c_H
                      \end{array}\right),
\label{eq:higgs}
\end{eqnarray}
in the $\bar{\nu}_H^c$, $\nu_H^c$ directions. The relevant
part of the superpotential, which includes the leading
non-renormalizable term, is
\begin{equation}
\delta W=\kappa S(-M^2+\bar{H}^c H^c)-
\beta\frac{S(\bar{H}^c H^c)^2}{M_S^2}~,
\label{eq:susyinfl}
\end{equation}
where $M_S\approx 5\times 10^{17}~{\rm GeV}$ is the string
scale and $\beta$ is taken positive for simplicity.
D-flatness implies that $\bar{H}^{c} \,^{*}=e^{i\theta}H^c$.
We restrict ourselves to the direction with $\theta=0$
($\bar{H}^{c} \,^{*}=H^c$) containing the `shifted'
inflationary path (see below). The scalar potential derived
from $\delta W$ then takes the form
\begin{equation}
V=\left[\kappa(\vert H^c\vert^2-M^2)-\beta\frac{\vert H^c
\vert^4}{M_S^2}\right]^2+2\kappa^2\vert S\vert^2
\vert H^c\vert^2
\left[1-\frac{2\beta}{\kappa M_S^2}\vert H^c\vert^2
\right]^2.
\label{eq:inflpot}
\end{equation}
Defining the dimensionless variables $w=\vert S\vert/M$,
$y=\vert H^c\vert/M$, we obtain
\begin{equation}
\tilde{V}=\frac{V}{\kappa^2M^4}=(y^2-1-\xi y^4)^2+
2w^2y^2(1-2\xi y^2)^2,
\label{eq:vtilde}
\end{equation}
where $\xi=\beta M^2/\kappa M_S^2$. This potential is a
simple extension of the standard potential for SUSY
\index{SUSY} hybrid inflation \index{hybrid inflation}
(which corresponds to $\xi=0$) and appears in a wide class
of models incorporating the leading non-renormalizable
correction to the standard hybrid inflationary superpotential.

\par
For constant $w$ (or $|S|$), $\tilde V$ in
(\ref{eq:vtilde}) has extrema at
\begin{equation}
y_1=0,~y_2=\frac{1}{\sqrt{2\xi}},~y_{3\pm}=\frac{1}
{\sqrt{2\xi}}\sqrt{(1-6\xi w^2)\pm\sqrt{(1-6\xi w^2)^2-
4\xi(1-w^2)}}.
\label{eq:extrema}
\end{equation}
Note that the first two extrema (at $y_1$, $y_2$) are
$|S|$-independent and, thus, correspond to classically flat
directions, the trivial one at $y_1=0$ with $\tilde{V}_1=1$,
and the `shifted' one at  $y_2=1/\sqrt{2\xi}={\rm constant}$
with $\tilde{V}_2=(1/4\xi-1)^2$, which we will use as
inflationary path. The trivial trajectory is a valley of minima
for $w>1$, while the `shifted' one for
$w>w_0=(1/8\xi-1/2)^{1/2}$, which is its critical point. We
take $\xi<1/4$, so that $w_0>0$ and the `shifted' path is
destabilized (in the chosen direction
$\bar{H}^{c} \,^{*}=H^c$) before $w$ reaches zero. The
extrema at $y_{3\pm}$, which are $|S|$-dependent and
non-flat, do not exist for all values of $w$ and $\xi$,
since the expressions under the square roots in
(\ref{eq:extrema}) are not always non-negative. These
two extrema, at $w=0$, become SUSY \index{SUSY} vacua. The
relevant SUSY \index{SUSY} vacuum (see below) corresponds to
$y_{3-}(w=0)$ and, thus, the common vev $v_0$ of
$\bar{H}^{c}$, $H^c$ is given by
\begin{equation}
\left(\frac{v_0}{M}\right)^2=
\frac{1}{2\xi}(1-\sqrt{1-4\xi})~.
\label{eq:v0}
\end{equation}

\par
We will now discuss the structure of $\tilde{V}$ and the
inflationary history for $1/6<\xi<1/4$. For fixed $w>1$,
there exist two local minima at $y_1=0$ and
$y_2=1/\sqrt{2\xi}$, which has lower potential energy
density, and a local maximum at $y_{3+}$ between
the minima. As $w$ becomes smaller than unity, the
extremum at $y_1$ turns into a local maximum, while the
extremum at $y_{3+}$ disappears. The system then falls
into the `shifted' path in case it had started at
$y_1=0$. As we further decrease $w$ below
$(2-\sqrt{36\xi-5})^{1/2}/3\sqrt{2\xi}$, a pair of
new extrema, a local minimum at $y_{3-}$ and a local
maximum at $y_{3+}$, are created between $y_1$ and
$y_2$. As $w$ crosses $(1/8\xi-1/2)^{1/2}$, the local
maximum at $y_{3+}$ crosses $y_2$ becoming a local
minimum. At the same time, the local minimum at $y_2$
turns into a local maximum and inflation
\index{inflation} ends with the system falling into
the local minimum at $y_{3-}$ which, at $w=0$, becomes
the SUSY \index{SUSY} vacuum.

\par
We see that, no matter where the system starts from, it
passes from the `shifted' path, where the relevant part
of inflation \index{inflation} takes place. So, $G_{PS}$
is broken during inflation \index{inflation} and no
monopoles \index{monopole} are produced at the `waterfall'.

\par
After inflation, \index{inflation} the system could fall
into the minimum at $y_{3+}$ instead of the one at $y_{3-}$.
This, however, does not happen since in the last e-folding or
so the barrier between the minima at $y_{3-}$ and $y_2$
is considerably reduced and the decay of the `false vacuum'
at $y_2$ to the minimum at $y_{3-}$ is completed within
a fraction of an e-folding before the $y_{3+}$ minimum
even appears. This transition is further accelerated by
the inflationary density perturbations.
\index{density perturbations}

\par
The mass spectrum on the `shifted' path can be evaluated
\cite{jean}. We find that the only mass splitting in
supermultiplets occurs in the $\bar{\nu}_H^c$,
$\nu_H^c$ sector. Namely, we obtain one Majorana fermion
with ${\rm mass}^2$ equal to $4\kappa^2\vert S\vert^2$,
which corresponds to the direction
$(\bar{\nu}_H^c+\nu_H^c)/\sqrt{2}$, and two normalized real
scalars $\mathrm{Re}(\delta\bar{\nu}^c_H+\delta\nu^c_H)$
and $\mathrm{Im}(\delta\bar{\nu}^c_H+\delta\nu^c_H)$ with
$m_{\pm}^2=4\kappa^2\vert S\vert^2\mp 2\kappa^2m^2$. Here,
$m=M(1/4\xi-1)^{1/2}$ and
$\delta\bar{\nu}^c_H=\bar{\nu}^c_H-v$,
$\delta\nu^c_H=\nu^c_H-v$ with
$v=(\kappa M_S^2/2\beta)^{1/2}$ being the common value of
$\bar{H}^c$, $H^c$ on the trajectory.

\par
The radiative corrections on the `shifted' path can then be
constructed using (\ref{eq:deltav}) and $(\delta T/T)_Q$
and $\kappa$ can be evaluated. We find the same expressions as
in (\ref{eq:qa}) and (\ref{eq:kappa}) with $N=2$ ($N=4$)
in the formula for $(\delta T/T)_Q$ ($\kappa$) and $M$
generally replaced by $m$. The COBE \index{COBE} \cite{cobe}
result can be reproduced, for instance, with
$\kappa\approx 4\times 10^{-3}$, which corresponds to
$\xi=1/5$, $v_0\approx 1.7\times 10^{16}~{\rm GeV}$ (we
put $N_Q\approx 55$, $\beta=1$). The scales
$M\approx 1.45\times 10^{16}~{\rm GeV}$,
$m\approx 7.23\times 10^{15}~{\rm GeV}$, the mass of the
inflaton \index{inflaton}
$m_{\mathrm{infl}}\approx 4.1\times 10^{13}~{\rm GeV}$ and
the `inflationary scale', which characterizes the inflationary
`vacuum' energy density, $v_{\rm infl}=\kappa^{1/2}m\approx
4.57\times 10^{14}~{\rm GeV}$. The spectral index
\index{spectral index} $n=0.954$.

\subsection{Smooth Hybrid Inflation
\index{smooth hybrid inflation}}
\label{subsec:smooth}

An alternative solution to the monopole \index{monopole}
problem of hybrid inflation \index{hybrid inflation} has been
proposed \cite{smooth} some years ago. We will present it here
within the SUSY \index{SUSY} PS model \index{PS model} of
Sect.\ref{subsec:shifted}, although it can be applied to other
semi-simple (and possibly some simple) gauge groups too. The
idea is to impose an extra $Z_2$ symmetry under which
$H^c\rightarrow -H^c$. The whole structure of the model
remains unchanged except that now only even powers of the
combination $\bar{H}^cH^c$ are allowed in the superpotential
terms.

\par
The inflationary superpotential in (\ref{eq:susyinfl})
becomes
\begin{equation}
\delta W=S\left(-\mu^2+\frac{(\bar{H}^cH^c)^2}
{M_S^2}\right),
\label{eq:smoothsuper}
\end{equation}
where we absorbed the dimensionless parameters $\kappa$,
$\beta$ in $\mu$, $M_S$. The resulting scalar potential
$V$ is then given by
\begin{equation}
\tilde{V}=\frac{V}{\mu^4}=(1-\tilde\chi^4)^2+
16\tilde\sigma^2\tilde\chi^6,
\label{eq:smoothpot}
\end{equation}
where we used the dimensionless fields $\tilde\chi=
\chi/2(\mu M_S)^{1/2}$, $\tilde\sigma=
\sigma/2(\mu M_S)^{1/2}$ with $\chi$, $\sigma$ being
normalized real scalar fields defined by
$\bar{\nu}_H^c=\nu_H^c=\chi/2$, $S=\sigma/\sqrt{2}$
after rotating $\bar{\nu}_H^c$, $\nu_H^c$, $S$ to the
real axis.

\par
The emerging picture is completely different. The flat
direction at $\tilde\chi=0$ is now a local maximum with
respect to $\tilde\chi$ for all values of $\tilde\sigma$,
and two new symmetric valleys of minima appear
\cite{smooth,shi} at
\begin{equation}
\tilde\chi=\pm\sqrt{6}\tilde\sigma\left[\left(1+
\frac{1}{36\tilde\sigma^4}\right)^{\frac{1}{2}}-1
\right]^{\frac{1}{2}}.
\label{eq:smoothvalley}
\end{equation}
They contain the SUSY \index{SUSY} vacua which lie at
$\tilde\chi=\pm 1$, $\tilde\sigma=0$. These valleys are
not classically flat. In fact, they possess a slope already at
the classical level, which can drive the inflaton
\index{inflaton} towards the vacua. Thus, there is no need of
radiative corrections in this case. The potential on these
paths is \cite{smooth,shi}
\begin{eqnarray}
\tilde{V}&=&48\tilde\sigma^4\left[72\tilde\sigma^4\left(1+
\frac{1}{36\tilde\sigma^4}\right)\left(\left(1+
\frac{1}{36\tilde\sigma^4}\right)^{\frac{1}{2}}-1\right)
-1\right]
\nonumber \\
&=&1-\frac{1}{216\tilde\sigma^4}+\cdots,~~{\rm for}
~\tilde\sigma\gg 1~.
\label{eq:smoothV}
\end{eqnarray}
The system follows, from the beginning, a particular
inflationary path and, thus, ends up at a particular
point of the vacuum manifold leading to no production of
disastrous monopoles. \index{monopole}

\par
The end of inflation \index{inflation} is not abrupt in this
case since
the inflationary path is stable with respect to $\tilde\chi$
for all $\tilde\sigma$'s. The value $\tilde\sigma_0$ of
$\tilde\sigma$ at which inflation \index{inflation} is
terminated smoothly is
found from the $\epsilon$ and $\eta$ criteria, and the
derivatives \cite{shi} of the potential on the inflationary
path:
\begin{equation}
\frac{d\tilde{V}}{d\tilde\sigma}=192\tilde\sigma^3
\left[(1+144\tilde\sigma^4)\left(\left(1+
\frac{1}{36\tilde\sigma^4}\right)^{\frac{1}{2}}
-1\right)-2\right],
\label{eq:firstder}
\end{equation}
\begin{eqnarray}
\frac{d^2\tilde{V}}{d\tilde\sigma^2}&=&
\frac{16}{3\tilde\sigma^2}
\Biggl\{(1+504\tilde\sigma^4)
\left[72\tilde\sigma^4\left(\left(1+
\frac{1}{36\tilde\sigma^4}\right)^{\frac{1}{2}}
-1\right)-1\right]
\nonumber \\
& &-(1+252\tilde\sigma^4)\left(\left(1+
\frac{1}{36\tilde\sigma^4}\right)^{-\frac{1}{2}}
-1\right)\Biggl\}.
\label{eq:secondder}
\end{eqnarray}

\par
The quantities $(\delta T/T)_Q$ and $N_Q$ can be found using
(\ref{eq:firstder}). One important advantage of this scenario
is that the common vev of $\bar{H}^c$, $H^c$, which is equal to
$v_0=(\mu M_S)^{1/2}$, is not so rigidly constrained and, thus,
can be chosen equal to the SUSY \index{SUSY} GUT \index{GUT}
scale ($v_0\approx 2.86\times 10^{16}~{\rm GeV})$. From COBE
\index{COBE} \cite{cobe} and for $N_Q\approx 57$, we then
obtain $M_S\approx 4.39\times 10^{17}~{\rm GeV}$ and
$\mu\approx 1.86\times 10^{15}~{\rm GeV}$, which are quite
`natural'. The value of $\sigma$ at which inflation
\index{inflation} ends corresponds to $\eta=-1$ and is
$\sigma_0\approx 1.34\times 10^{17}~{\rm GeV}$. The value of
$\sigma$ at which our present horizon \index{horizon} crosses
outside the inflationary horizon \index{horizon} is
$\sigma_Q \approx 2.71 \times 10^{17}~{\rm GeV}$. The
inflaton \index{inflaton} mass is
$m_{\rm infl}=2\sqrt{2}\mu^2/v_0\approx
3.42\times 10^{14}~{\rm GeV}$.

\section{`Reheating' \index{reheating} and the Gravitino
Constraint \index{gravitino constraint}}
\label{sec:reheat}

A complete inflationary scenario should be followed by a
successful `reheating' \index{reheating} which satisfies
the gravitino constraint \index{gravitino constraint}
\cite{gravitino} and generates the
observed BAU. \index{BAU} We will discuss `reheating'
\index{reheating} within a SUSY \index{SUSY} GUT
\index{GUT} leading to standard hybrid inflation.
\index{hybrid inflation} We consider a moderate extension
of the minimal supersymmetric standard model (MSSM) based
on the left-right symmetric gauge group
\begin{equation}
G_{LR}=SU(3)_c\times SU(2)_L\times SU(2)_R
\times U(1)_{B-L}
\label{eq:lr}
\end{equation}
(see \cite{lss,hier,trieste}). The breaking of
$G_{LR}$ to $G_S$ is achieved via a conjugate pair of
$SU(2)_R$ doublets $\bar l^{c}$, $l^c$ with $B-L$
(baryon minus lepton number) equal to -1, 1, which
acquire vevs along their right handed neutrino directions
$\bar\nu^{c}_{H}$, $\nu^c_H$ corresponding to
$\bar{\phi}$, $\phi$ in Sect.\ref{subsec:susy}.
The relevant superpotential is
\begin{equation}
W=\kappa S(-M^2+\bar l^{c}l^c),
\label{eq:W}
\end{equation}
where $\kappa$, $M$ are made positive by field redefinitions.
This superpotential leads to hybrid inflation
\index{hybrid inflation} exactly as $W$
in (\ref{eq:superpot}). $(\delta T/T)_Q$ and $\kappa$
are given by (\ref{eq:qa}) and (\ref{eq:kappa}) with
$N=2$ since $\bar l^{c}$, $l^c$ have two components each.

\par
$G_{LR}$ implies the presence of right handed neutrino
superfields $\nu^c_i$ (with $i=1,2,3$), which form $SU(2)_R$
doublets $L^c_i=(\nu^c_i,e^c_i)$ with the $SU(2)_L$ singlet
charged antileptons $e^c_i$. Intermediate scale $\nu^c$
masses are generated via the superpotential terms
$\gamma_i\bar l^{c}\bar l^{c}L^c_iL^c_i/m_P$ (in a basis
with diagonal and positive $\gamma$'s). These masses are
$M_i=2\gamma_iM^2/m_P$
($\langle\bar l^{c}\rangle$, $\langle l^{c}\rangle>0$
by a $B-L$ rotation). Light neutrinos acquire hierarchical
seesaw masses and, thus, cannot play the role of hot dark
matter (HDM) in the universe. (This requires degenerate
neutrino masses which can be obtained \cite{trieste,deg} via
$SU(2)_L$ triplets \cite{triplet}.) They are suitable for a
universe with non-zero cosmological constant favored by
recent observations \cite{lambda}. In this case, HDM is not
necessary \cite{lambdafit,primack}. The terms generating the
$\nu^c$ masses also cause the decay of the inflaton
\index{inflaton} (see Sect.\ref{sec:reheat}).

\par
After the end of inflation, \index{inflation} the system
falls towards the SUSY \index{SUSY} vacuum and performs damped
oscillations about it. The inflaton \index{inflaton}
(oscillating system) consists of the two
complex scalar fields $\theta=(\delta\bar\nu^c_H+
\delta\nu^c_H)/\sqrt{2}$ ($\delta\bar\nu^c_H=
\bar\nu^c_H-M$, $\delta\nu^c_H=\nu^c_H-M$) and $S$,
with equal mass $m_{\rm infl}=\sqrt{2}\kappa M$.

\par
The oscillating fields $\theta$ and $S$ decay into a pair of
right handed neutrinos ($\psi_{\nu^c_i}$) and sneutrinos
($\nu^c_i$) respectively via the superpotential
couplings $\bar{l}^c\bar{l}^c L^cL^c$ and $S\bar{l}^cl^c$.
The relevant Lagrangian terms are:
\begin{equation}
L^\theta_{\rm decay}=-\sqrt{2}\gamma_i\frac{M}{m_P}
\theta\psi_{\nu_i^c}\psi_{\nu_i^c}+ h.c.~,
\label{eq:thetadecay}
\end{equation}
\begin{equation}
L^S_{\rm decay}=-\sqrt{2}\gamma_i\frac{M}{m_P}S^*\nu^c_i
\nu^c_im_{\rm infl}+h.c.~,
\label{eq:sdecay}
\end{equation}
and the common, as it turns out, decay width is given by
\begin{equation}
\Gamma=\Gamma_{\theta\rightarrow\bar\psi_{\nu^c_i}
\bar\psi_{\nu^c_i}}=\Gamma_{S\rightarrow\nu^c_i\nu^c_i}=
\frac{1}{8\pi}\left(\frac{M_i}{M}\right)^2m_{\rm infl}~,
\label{eq:gammainfl}
\end{equation}
provided that the relevant $\nu^c$ mass $M_i<m_{\rm infl}/2$.

\par
To minimize the number of small coupling constants, we assume
that
\begin{equation}
M_2<\frac{1}{2}m_{\rm infl}\leq M_3=\frac{2M^2}{m_P}~~
({\rm with}~\gamma_3=1)~,
\label{eq:ineq}
\end{equation}
so that the inflaton \index{inflaton} decays into the second
heaviest right handed neutrino superfield $\nu^c_2$ with mass
$M_2$. The second inequality in (\ref{eq:ineq}) implies that
$y_Q\leq\sqrt{2N_Q}/\pi\approx 3.34$ for $N_Q\approx 55$.
This gives $x_Q\stackrel{_{<}}{_{\sim}}3.5$. As an
example, choose $x_Q\approx 1.05$ (bigger values cannot
give adequate BAU) \index{BAU} which yields
$y_Q\approx 0.28$. From the COBE \index{COBE} \cite{cobe}
result, we then obtain $M\approx 4.06\times
10^{15}~{\rm GeV}$, $\kappa\approx 4\times 10^{-4}$,
$m_{\rm infl}\approx 2.3\times 10^{12}~{\rm GeV}$ and
$M_3\approx 1.35\times 10^{13}~{\rm GeV}$.

\par
The `reheat' temperature \index{`reheat' temperature} $T_r$,
for the MSSM spectrum, is given by \cite{lss}
\begin{equation}
T_r\approx\frac{1}{7}(\Gamma M_P)^{\frac{1}{2}},
\label{eq:reheatmssm}
\end{equation}
and must satisfy the gravitino constraint
\index{gravitino constraint} \cite{gravitino},
$T_r\stackrel{_{<}}{_{\sim}}10^9~{\rm GeV}$, for
gravity-mediated SUSY \index{SUSY} breaking with universal
boundary conditions. To maximize the `naturalness' of the
model, we take the maximal $M_2$ (and, thus, $\gamma_2$)
allowed by this constraint. This is
$M_2\approx 2.7\times 10^{10}~{\rm GeV}$ ($\gamma_2
\approx 2\times 10^{-3}$). Note that, with this $M_2$,
the first inequality in (\ref{eq:ineq}) is well
satisfied.

\section{Baryogenesis \index{baryogenesis} via
Leptogenesis \index{leptogenesis}}
\label{sec:leptogenesis}

\subsection{Primordial Leptogenesis \index{leptogenesis}}
\label{subsec:primordial}

In hybrid inflationary models, it is \cite{lepto} generally
not so convenient to generate the observed BAU \index{BAU}
in the usual way, i.e., through the decay of superheavy color
(anti)triplets. Some of the reasons are:
\begin{list}
\setlength{\rightmargin=0cm}{\leftmargin=0cm}
\item[{\bf i.}] $B$ is practically conserved in most models
of this type. In some cases \cite{bparity}, this is due to
a discrete `baryon parity' symmetry. In the left-right model
under consideration, $B$ is exactly conserved because of a
$U(1)_R$.
\vspace{.25cm}
\item[{\bf ii.}] The gravitino constraint
\index{gravitino constraint} would require that the mass of
the (anti)triplets does not exceed $10^{10}~{\rm GeV}$. This
suggests strong deviations from the MSSM gauge coupling
unification and possibly proton instability.
\end{list}

\par
It is generally preferable to produce an initial lepton
asymmetry \cite{leptogenesis} \index{lepton asymmetry}
which is then partly turned into baryon asymmetry
\index{baryon asymmetry} by sphalerons \index{sphaleron}
\cite{dimopoulos,sphaleron}. In the left-right model we
consider and in many other models, this is the only way
for obtaining the BAU \index{BAU} since the inflaton
\index{inflaton} decays into right handed neutrino
superfields. Their subsequent decay to lepton (antilepton)
$L$ ($\bar{L}$) and electroweak Higgs superfields can
only produce a lepton asymmetry. \index{lepton asymmetry}
It is important to ensure that this asymmetry is not erased
\cite{turner} by lepton number violating $2 \rightarrow 2$
scattering processes such as
$LL\rightarrow h^{(1)}\,^{*}h^{(1)}\,^{*}$ or
$Lh^{(1)}\rightarrow \bar{L}h^{(1)}\,^{*}$ at all
$T$'s between $T_{r}$ and $100~{\rm GeV}$ ($h^{(1)}$
is the Higgs $SU(2)_L$ doublet which couples to up type
quarks). This is satisfied since the lepton asymmetry
\index{lepton asymmetry} is protected \cite{ibanez} by
SUSY \index{SUSY} at $T$'s between $T_r$ and
$T\sim 10^{7}~{\rm GeV}$ and, for
$T\stackrel{_{<}}{_{\sim }}10^{7}~{\rm GeV}$, these
scattering processes are well out of equilibrium provided
\cite{ibanez}
$m_{\nu_{\tau}}\stackrel{_<}{_\sim} 10~{\rm{eV}}$.
For MSSM spectrum, the observed BAU \index{BAU} $n_B/s$
is related \cite{ibanez} to the primordial lepton asymmetry
\index{lepton asymmetry} $n_L/s$ by
$n_B/s=(-28/79)n_L/s$ (see Sect.\ref{subsec:sphaleron}).

\par
The lepton asymmetry \index{lepton asymmetry} is generated
via the decay of the superfield $\nu^{c}_{2}$, produced by
the inflaton \index{inflaton} decay, to electroweak Higgs
and (anti)lepton superfields. The relevant one-loop diagrams
are both of the vertex and self-energy type \cite{covi} with
an exchange of $\nu^{c}_{3}$. The resulting lepton asymmetry
\index{lepton asymmetry} is
\cite{neu}
\begin{equation}
\frac{n_{L}}{s}\approx 1.33~\frac{9T_{r}}
{16\pi m_{\rm infl}}~\frac{M_2}{M_3}
~\frac{{\rm c}^{2}{\rm s}^{2}\sin 2\delta
(m_{3}^{D}\,^{2}-m_{2}^{D}\,^{2})^{2}}
{|\langle h^{(1)}\rangle|^{2}~(m_{3}^{D}\,^{2}
{\rm \ s}^{2}+m_{2}^{D}\,^{2}{\rm \ c^{2}})}~,
\label{eq:leptonasym}
\end{equation}
where $|\langle h^{(1)}\rangle|\approx 174~\rm{GeV}$
(for large $\tan\beta$), $m_{2,3}^{D}$ are the `Dirac'
masses of the neutrinos (in a basis where they are diagonal
and positive), and ${\rm c}=\cos\theta$,
${\rm s}=\sin\theta$, with $\theta$ and $\delta$ being
the rotation angle and phase which diagonalize the Majorana
mass matrix of the $\nu^c$'s. Equation (\ref{eq:leptonasym})
holds \cite{pilaftsis} provided that $M_{2}\ll M_{3}$ and
the decay width of $\nu^{c}_{3}$ is
$\ll(M_{3}^{2}-M_{2}^{2})/M_{2}$, which are satisfied in
our model. Here, we considered only the two heaviest families
($i=2,3$) ignoring the first one since the analysis
\cite{giunti} of the CHOOZ experiment \cite{chooz} has shown
that the solar and atmospheric neutrino oscillations
\index{neutrino oscillations} decouple.

\par
The light neutrino mass matrix is given by the seesaw
formula:
\begin{equation}
m_{\nu}\approx-\tilde m^{D}\frac{1}{M}m^{D},
\label{eq:neumass}
\end{equation}
where $m^{D}$ is the `Dirac' neutrino mass matrix and
$M$ the Majorana $\nu^c$ mass matrix. The determinant
and the trace invariance of the light neutrino mass
matrix imply \cite{neu} two constraints on the
(asymptotic) parameters:
\begin{equation}
m_{2}m_{3}=\frac{\left(m_{2}^{D}m_{3}^{D}\right)^{2}}
{M_{2}M_{3}}~,
\label{eq:determinant}
\end{equation}
\begin{eqnarray*}
m_{2}\,^{2}+m_{3}\,^{2}=\frac{\left(m_{2}^{D}\,^{2}
{\rm \ c}^{2}+m_{3}^{D}\,^{2}{\rm \ s}^{2}\right)^{2}}
{M_{2}\,^{2}}+
\end{eqnarray*}
\begin{equation}
\frac{\left(m_{3}^{D}\,^{2}{\rm \ c}^{2}+
m_{2}^{D}\,^{2}{\rm \ s}^{2}\right)^{2}}{M_{3}\,^{2}}+
\frac{2(m_{3}^{D}\,^{2}-m_{2}^{D}\,^{2})^{2}
{\rm c}^{2}{\rm s}^{2}{\cos 2\delta }}{M_{2}M_{3}}~,
\label{eq:trace}
\end{equation}
where $m_{2}=m_{\nu_\mu}$ and $m_{3}=m_{\nu_\tau}$ are the
(positive) eigenvalues of $m_{\nu}$.

\par
The $\mu-\tau$ mixing angle $\theta_{23}=\theta _{\mu\tau}$
lies \cite{neu} in the range
\begin{equation}
|\,\varphi -\theta ^{D}|\leq \theta _{\mu\tau}\leq
\varphi +\theta^{D},\ {\rm {for}\ \varphi +
\theta }^{D}\leq \ \pi /2~,
\label{eq:mixing}
\end{equation}
where $\varphi$ is the rotation angle which diagonalizes
the light neutrino mass matrix in the basis where the `Dirac'
mass matrix is diagonal and $\theta ^{D}$ is the `Dirac'
mixing angle, i.e., the `unphysical' mixing angle with zero
Majorana  masses of the right handed neutrinos.

\par
We take $m_{\nu_{\mu}}\approx 2.6\times 10^{-3}~\rm{eV}$
and $m_{\nu _{\tau }}\approx 7\times 10^{-2}~\rm{eV}$
which are the central values from the small angle MSW
\index{MSW} resolution of the solar neutrino problem
\cite{smirnov} and SuperKamiokande \index{SuperKamiokande}
\cite{superk}. We choose $\delta\approx-\pi/4$ to maximize
$-n_L/s$. Finally, we assume that $\theta^D\approx 0$, so
that maximal $\nu_\mu-\nu_\tau$ mixing, which is favored by
SuperKamiokande \index{SuperKamiokande} \cite{superk},
corresponds to $\varphi\approx\pi/4$.

\par
From (\ref{eq:determinant}) and (\ref{eq:trace}) and
the diagonalization of $m_\nu$, we determine the value of
$m_{3}^{D}$ corresponding to $\varphi\approx\pi/4$ for
any given $\kappa$. A solution for $m_{3}^{D}$ exists
provided that $M_2\stackrel{_{<}}{_{\sim }}0.037M_3$.
For the numerical example in Sect.\ref{sec:reheat}, we find
$m_{3}^{D}\approx 8.3~{\rm GeV}$, $m_{2}^{D}
\approx 0.98~{\rm GeV}$ and
$n_L/s\approx -2.23\times 10^{-10}$, which satisfies the
baryogenesis \index{baryogenesis} constraint. Thus, with
`natural' values of $\kappa$ ($\approx 4\times 10^{-4}$)
and the other relevant parameters
($\gamma_2\approx 2\times 10^{-3}$,
$\gamma_3\approx 1$), we were able not only to reproduce
COBE \index{COBE} \cite{cobe} but also to have a successful
`reheating' \index{reheating} satisfying the gravitino and
baryogenesis \index{baryogenesis} constraints
\index{gravitino constraint} together with the requirements
from solar and atmospheric neutrino oscillations.
\index{neutrino oscillations} Similar results hold
\cite{jean,shi} for the shifted and smooth hybrid
inflationary models of Sect.\ref{sec:extensions}.

\subsection {Sphaleron \index{sphaleron} Effects}
\label{subsec:sphaleron}

\par
To see how the lepton asymmetry \index{lepton asymmetry}
partly turns into baryon asymmetry, \index{baryon asymmetry}
we must first discuss the non-perturbative
baryon and lepton number violation \cite{thooft} in the
standard model. Consider the electroweak gauge symmetry
$SU(2)_{L}\times U(1)_{Y}$ in the limit where the
Weinberg angle $\theta_W=0$ and concentrate on
$SU(2)_{L}$ ($\theta_{W}\neq 0$ does not alter the
conclusions). Also, for the moment, ignore the fermions
and Higgs fields so as to have a pure $SU(2)_{L}$ gauge
theory. This theory has \cite{vacuum}
infinitely many classical vacua which are topologically
distinct and are characterized by a `winding number'
$n\in Z$. In the `temporal gauge' ($A_{0}=0$), the
remaining gauge freedom consists of time independent
transformations and the vacuum is a pure gauge
\begin{equation}
A_{i} = \frac{i}{g}~\partial _{i}g(\bar{x})
g ^{-1}(\bar{x})~,~i=1,2,3~.
\label{eq:gauge}
\end{equation}
Here $g$ is the $SU(2)_{L}$ gauge coupling constant,
$\bar{x}\in$3-space, $g(\bar{x})\in SU(2)_{L}$, and
$g(\bar{x})\rightarrow 1$ as
$\mid\bar{x}\mid\rightarrow\infty$. Thus, the 3-space
compactifies to a sphere $S^{3}$ and $g(\bar{x})$ gives
a map: $S^{3} \rightarrow SU(2)_{L}$ ($SU(2)_{L}$ is
topologically equivalent to $S^{3}$). These maps are
classified into homotopy classes constituting the third
homotopy group of $S^{3},~\pi_{3}(S^{3})$, and are
characterized by a `winding number'
\begin{equation}
n=\int d^{3}x~\epsilon^{ijk}~{\rm{tr}}
\left(\partial_{i}g(\bar{x})
g^{-1}(\bar{x})\partial_{j}g(\bar{x}) g^{-1}(\bar{x})
\partial_{k}g(\bar{x}) g^{-1}(\bar{x})\right).
\label{eq:wind}
\end{equation}
The corresponding vacua are denoted as $\mid n\rangle$,
$n\in Z$.

\par
The tunneling amplitude from the vacuum $\mid n_{-}
\rangle$ at $t=-\infty$ to the vacuum $\mid n_{+}
\rangle~$ at $t=+\infty$ is given by the functional
integral
\begin{equation}
\langle n_{+}\mid n_{-}\rangle=\int(dA)~e^{-S(A)}
\label{eq:path}
\end{equation}
over all gauge field configurations satisfying the
appropriate boundary conditions at $t=\pm \infty$.
Performing a Wick rotation,
$x_0\equiv t\rightarrow -i x_{4}$,
we go to Euclidean space-time. Any Euclidean field
configuration with finite action is characterized by an
integer known as the Pontryagin number
\begin{equation}
q=\frac {g^{2}}{16\pi^{2}}
\int d^{4}x~{\rm{tr}}\left(F^{\mu \nu}
\tilde{F}_{\mu\nu}\right),
\label{eq:pontryagin}
\end{equation}
where $\mu$,$\nu$=1,2,3,4 and $\tilde{F}_{\mu \nu}=
\frac {1}{2}\epsilon_{\mu \nu \lambda \rho}
F^{\lambda \rho}$ is the  dual field strength. It is
known that ${\rm{tr}}(F^{\mu\nu}\tilde{F}_{\mu\nu})=
\partial^{\mu}J_{\mu}$, where $J_{\mu}$ is the
`Chern-Simons current' given by
\begin{equation}
J_{\mu}=\epsilon_{\mu\nu\alpha\beta}~{\rm{tr}}
\left(A^{\nu}
F^{\alpha \beta}-\frac{2}{3}gA^{\nu}A^{\alpha}
A^{\beta}\right).
\label{eq:csc}
\end{equation}
In the `temporal gauge' ($A_0=0$),
\begin{eqnarray*}
q=\frac{g^{2}}{16 \pi^{2}} \int d^{4}x
~\partial^{\mu}J_{\mu}=
\frac{g^{2}}{16\pi^{2}}
\mathop{\Delta}_{x_{4}=\pm \infty}
\int d^{3}x~J_{0}
\end{eqnarray*}
\begin{equation}
=\frac{1}{24\pi^{2}}\mathop{\Delta}_{x_{4}=
\pm \infty}
\int d^{3}x~\epsilon^{ijk}~
{\rm{tr}}\left(\partial_{i} g g^{-1}\partial_{j}
gg^{-1}\partial_{k}gg^{-1}\right)=n_{+}-n_{-}~.
\label{eq:interpol}
\end{equation}
Thus, the Euclidean field configurations which
interpolate between the vacua $\mid n_{+}\rangle$,
$\mid n_{-}\rangle$ at $x_4=\pm\infty$ have
Pontryagin number $q=n_{+}-n_{-}$ and the  path integral
in (\ref{eq:path}) should be performed over all these
configurations.

\par
For a given $q$, there is a lower bound on $S(A)$,
\begin{equation}
S(A)\geq\frac{8 \pi^{2}}{g^{2}}\mid q\mid~,
\label{eq:lbound}
\end{equation}
which is saturated if and only if $F_{\mu\nu}=
\pm\tilde{F}_{\mu\nu}$, i.e, if the configuration
is self-dual or self-antidual. For $q$=1, the self-dual
classical solution is called instanton \cite{instanton}
\index{instanton} and is given by (in the `singular'
gauge)
\begin{equation}
A_{a \mu}(x)=\frac{2 \rho^{2}}{g(x-z)^{2}}
~\frac{\eta_{a \mu \nu} (x-z)^{\nu}}
{(x-z)^{2}+\rho^{2}}~,
\label{eq:instanton}
\end{equation}
where $\eta_{a\mu\nu}$ ($a$=1,2,3; $\mu$,$\nu$=
1,2,3,4) are the t' Hooft symbols with $\eta_{aij}=
\epsilon_{aij}$ ($i$,$j$=1,2,3), $\eta_{a4i}=
-\delta_{ai}$, $\eta_{ai4}=\delta_{ai}$ and
$\eta_{a44}=0$. The instanton \index{instanton}
depends on four Euclidean coordinates $z_{\mu}$ (its
position) and its scale (or size) $\rho$. Two
successive vacua $\mid n\rangle$, $\mid n+1\rangle$
are separated by a potential barrier of height
$\propto\rho^{-1}$. The Euclidean action of the
interpolating instanton \index{instanton} is always
equal to $8\pi^{2}/g^{2}$, but the height of the
barrier can be made arbitrarily small since the size
$\rho$ of the instanton \index{instanton} can be
taken arbitrarily large.

\par
We now reintroduce the fermions into the theory and
observe \cite{thooft} that the baryon and lepton number
currents carry anomalies, i.e.,
\begin{equation}
\partial_{\mu}J^{\mu}_{B}=
\partial _{\mu}J^{\mu}_{L}=
-n _{g}\frac{g^{2}}{16 \pi ^{2}}~{\rm{tr}}
(F_{\mu \nu}\tilde{F}^{\mu \nu})~,
\label{eq:anomaly}
\end{equation}
where $n_{g}$ is the number of generations. Consequently,
the tunneling from $\mid n_{-}\rangle$ to
$\mid n_{+}\rangle$ is accompanied by a change of the
baryon and lepton numbers $\Delta B=\Delta L=-n_{g}q=
-n_{g}(n_{+}-n_{-})$. We should note that
i) $\Delta (B-L)=0$, and ii) for $q$=1,
$\Delta B=\Delta L=-3$ which means that one lepton per
family and one quark per family and color are annihilated
(12-point function).

\par
We, finally, reintroduce the electroweak Higgs doublet
$h$ whose vev is
\begin{equation}
<h>=\frac {v}{\sqrt{2}}~\left(\matrix{0\cr 1\cr}
\right),~v\approx 246~{\rm{GeV}}.
\label{eq:vev}
\end{equation}
The instanton \index{instanton} then ceases to exist as
an exact solution. It is replaced by the so-called
`restricted instanton' \cite{restricted}
\index{restricted instanton} which is an approximate
solution for $\rho\ll v^{-1}$. For
$\mid x-z\mid\ll\rho$, the gauge field of the
`restricted instanton' \index{restricted instanton}
essentially coincides with that of the instanton
\index{instanton} and the Higgs field is
\begin{equation}
h(x)\approx\frac{v}{\sqrt{2}}
~\left(\frac{(x-z)^{2}}
{(x-z)^{2}+\rho^{2}}\right)^{1/2}
\left(\matrix {0\cr 1\cr}\right).
\label{eq:restricted}
\end{equation}
For $\mid x-z\mid\gg\rho$, the gauge and Higgs fields
decay to a pure gauge and the vev in (\ref{eq:vev})
respectively. The action of the `restricted instanton'
\index{restricted instanton} is
$S_{ri}=(8 \pi^{2}/g^{2})+\pi^{2} v^{2} \rho^{2}+
\cdots$, and thus the contribution of big size
`restricted instantons' \index{restricted instanton} to
the path integral in (\ref{eq:path}) is suppressed. This
justifies {\it a posteriori} the fact that we restricted
ourselves to solutions with $\rho\ll v^{-1}$.

\par
The height of the potential barrier between the vacua
$\mid n\rangle$, $\mid n+1\rangle$ cannot be now
arbitrarily small. Indeed, the static energy of the
`restricted instanton' \index{restricted instanton} at
$x_{4}=z_{4}$ ($\lambda$ is the Higgs self-coupling),
\begin{equation}
E_{b}(\rho)\approx\frac{3\pi^{2}}
{g^{2}}~\frac{1}{\rho}+
\frac{3}{8}\pi^{2}v^{2}\rho^{2}+
\frac{\lambda}{4}\pi^{2}v^{4}\rho^{3},
\label{eq:static}
\end{equation}
is minimized for
\begin{equation}
\rho_{{\rm{min}}}=\frac{\sqrt{2}}{gv}
\left(\frac{\lambda}
{g^{2}}\right)^{-1/2}\left(\left(\frac{1}{64}+
\frac{\lambda}{g^{2}}\right)^{1/2}-
\frac{1}{8}\right)^{1/2}
\sim M^{-1}_{W},
\label{eq:rhomin}
\end{equation}
and, thus, the minimal height of the potential barrier
is $E_{{\rm{min}}}\sim M_{W}/\alpha_W$
($\alpha_{W}=g^{2}/4 \pi$). The static solution
which corresponds to the top (saddle point) of this
potential barrier is called sphaleron \cite{sphaleronsol}
\index{sphaleron} and is given by
\begin{equation}
\bar{A}=v~\frac{f(\xi)}{\xi}~\hat{r}
\times\bar{\tau}~,
~h=\frac{v}{\sqrt{2}}~t(\xi)~\hat{r}\cdot
\bar{\tau}\left(\matrix{0\cr 1\cr}\right),
\label{eq:sphaleron}
\end{equation}
where $\xi=2M_{W}r$, $\hat{r}$ is the  radial unit
vector in 3-space
and the 3-vector $\bar{\tau}$ consists of the Pauli
matrices. The functions $f(\xi)$, $t(\xi)$, which can be
determined numerically, tend to zero as $\xi\rightarrow 0$
and to 1 as $\xi\rightarrow\infty$. The mass (static
energy) of the sphaleron \index{sphaleron} solution is
estimated to be
\begin{equation}
E_{{\rm{sph}}}=\frac{2M_{W}}{\alpha_{W}}~k~,
~1.5\leq k\leq 2.7~,~{\rm{for}}~0\leq
\lambda\leq\infty~,
\label{eq:sphmass}
\end{equation}
and lies between 10 and 15 TeV.

\par
At $T=0$ the tunneling from $\mid n\rangle$ to
$\mid n+1\rangle$ is utterly suppressed \cite{thooft} by
the factor ${\rm exp}(-8\pi^{2}/g^{2})$. At high $T$'s,
however, thermal fluctuations over the potential barrier are
frequent and the tunneling rate is \cite{dimopoulos,sphaleron}
enhanced. For $M_{W} \stackrel{_{<}}{_{\sim }}
T\stackrel{_{<}}{_{\sim }}T_{c}$ ($T_c$ is the
critical temperature of the electroweak transition), this
rate is calculated \cite{sphaleron} by expanding around
the sphaleron. \index{sphaleron} We find
\begin{equation}
\Gamma \approx 10^{4}~ n_{g}~ \frac {v (T)^{9}}{T^{8}}
~{\rm{exp}} (-E_{{\rm{sph}}}(T)/T)~.
\label{eq:sphrate}
\end{equation}
For a second order electroweak transition, $v(T)$,
$E_{{\rm{sph}}}(T)\propto (1-T^{2}/T^{2}_{c})^{1/2}$.
We can then show that $\Gamma\gg H$ for $T$'s between
$\sim 200~{\rm GeV}$ and $\sim T_{c}$. Furthermore, for
$T\geq T_{c}$, where the sphaleron \index{sphaleron}
ceases to exist, it was argued \cite{dimopoulos,sphaleron}
that we still have $\Gamma\gg H$. The overall conclusion
is that non-perturbative baryon and lepton number violating
processes are in equilibrium in the universe for
$T\stackrel{_{>}}{_{\sim }}200~{\rm GeV}$. Note that
$B-L$ is conserved by these processes.

\par
Given a primordial lepton asymmetry,
\index{lepton asymmetry} one can calculate
\cite{turner,ibanez} the resulting $n_{B}/s$. In MSSM,
the $SU(2)_{L}$ instantons \index{instanton} produce the
effective operator
\begin{equation}
O_{2}=(q q q l)^{n_{g}}(\tilde{h}^{(1)}
\tilde{h}^{(2)})\tilde{W}^{4},
\label{eq:woperator}
\end{equation}
and the $SU(3)_{c}$ instantons \index{instanton} the
operator
\begin{equation}
O_3=( q q u^{c}d^{c})^{n_{g}}\tilde{g}^{6},
\label{eq:coperator}
\end{equation}
where $q$ and $l$ are the quark and lepton $SU(2)_{L}$
doublets, $u^{c}$ and $d^{c}$ the up and down type
antiquark $SU(2)_{L}$ singlets, $h^{(2)}$ the Higgs
$SU(2)_{L}$ doublet which couples to down type quarks,
$g$ and $W$ the gluons and $W$ bosons, and tilde denotes
the superpartner. These interactions as well as the usual
MSSM interactions are in equilibrium at high $T$'s. The
equilibrium number density of an ultrarelativistic
particle species
$\Delta n\equiv n_{\rm part}-n_{\rm antipart}$ is
given by
\begin{equation}
\frac{\Delta n}{s}=\frac{15 g}{4\pi^{2}g_{*}}
\left(\frac{\mu}{T}\right)\epsilon~,
\label{eq:chemical}
\end{equation}
where $g$ is the number of internal degrees of freedom of
the particle, $\mu$ its chemical potential and $\epsilon=2$
or 1 for bosons or fermions. For each interaction in
equilibrium, the algebraic sum of the $\mu$'s of the
particles involved is zero. These constraints leave only two
independent chemical potentials, $\mu_{q} $ and
$\mu_{\tilde{g}}$. The baryon and lepton asymmetries are
then expressed \cite{ibanez} as
$$
\frac{n_{B}}{s}=\frac{30}{4 \pi^{2}g_{*}T}
(6n_{g} \mu_{q}-(4n_{g}-9)\mu_{\tilde{g}})~,
$$
\begin{equation}
\frac{n_{L}}{s}=-\frac{45}{4\pi^{2}g_{*}T}
\left(\frac{n_{g}(14n_{g} +9)}
{1+2n_{g}}\mu_{q}+\Omega(n_{g})
\mu_{\tilde{g}}\right),
\label{eq:bla}
\end{equation}
where $\Omega(n_{g})$ is a known \cite{ibanez} function.
Soft SUSY \index{SUSY} breaking couplings come in equilibrium
at $T \stackrel{_{<}}{_{\sim }}10^7~{\rm GeV}$
since their rate $\Gamma_{S}\approx m^{2}_{3/2}/T
\stackrel{_{>}}{_{\sim }}H\approx 30~T^{2}/M_{P}$.
In particular, the non-vanishing gaugino mass implies
$\mu _{\tilde{g}}=0$. Equation (\ref{eq:bla}) then
gives
\cite{ibanez}
\begin{equation}
\frac {n_{B}}{s} = \frac {4(1+2n_{g})}{22n_{g}+13}
~\frac {n_{B-L}}{s}~.
\label{eq:bbminl}
\end{equation}
Equating $n_{B-L}/s$ with the primordial $n_{L}/s$, we
get $n_{B}/s=(-28/79)n_{L}/s$, for $n_{g}=3$. Note
that it is crucial to generate a primordial $n_{B-L}/s$
and not only a $n_{B}/s$ (and $n_{L}/s$) since
otherwise the final $n_{B}/s$ will vanish. This is
another reason which disfavors the creation of the BAU
\index{BAU} via the decay of superheavy color
(anti)triplets since their interactions usually conserve
$B-L$.

\section{Conclusions}
\label{sec:conclusions}

\par
We have summarized the shortcomings of the SBB \index{SBB}
model. We have then shown how they are resolved by inflationary
cosmology which suggests that the universe, in its early stages,
underwent a period of exponential expansion driven by an almost
constant `vacuum' energy density. This may have happened during
the GUT \index{GUT} phase transition at which the Higgs field
which breaks the GUT \index{GUT} gauge symmetry was displaced
from the vacuum. This field (inflaton) \index{inflaton} could
then, for some time, roll slowly towards the vacuum providing
the `vacuum' energy density. Inflation \index{inflation}
generates the primordial density perturbations
\index{density perturbations} which are necessary for
the large scale structure formation \index{structure formation}
in the universe and the observed temperature fluctuations
\index{temperature fluctuations} of the CMBR. \index{CMBR}
After the end of inflation, \index{inflation} the inflaton
\index{inflaton} performs damped oscillations about the
vacuum and eventually decays into light particles `reheating'
\index{reheating} the universe.

\par
The early realizations of inflation \index{inflation} required
`unnaturally' small coupling constants. This problem was solved
by the so-called hybrid inflationary scenario which uses two real
scalar fields instead of one that was customarily used. One of
them provides the `vacuum' energy density for inflation
\index{inflation} while the other one is the slowly rolling
field. Hybrid inflation \index{hybrid inflation} arises
`naturally' in many SUSY \index{SUSY} GUTs. \index{GUT}
However, the cosmological disaster from the overproduction of
GUT \index{GUT} monopoles, \index{monopole} which was avoided
in earlier inflationary models, reappears in hybrid inflation.
\index{hybrid inflation} We have constructed two `natural'
extensions of SUSY \index{SUSY} hybrid inflation
\index{hybrid inflation} which do not suffer from the monopole
\index{monopole} problem.

\par
We have shown that successful `reheating' \index{reheating}
satisfying the gravitino constraint \index{gravitino constraint}
on the `reheat' temperature \index{`reheat' temperature} takes
place after the end of inflation \index{inflation} in all three
versions of hybrid inflation \index{hybrid inflation} we have
considered here. Adequate baryogenesis \index{baryogenesis} via
a primordial leptogenesis \index{leptogenesis} occurs
consistently with the solar and atmospheric neutrino oscillation
\index{neutrino oscillations} data. The primordial lepton
asymmetry \index{lepton asymmetry} is turned partly into baryon
asymmetry \index{baryon asymmetry} via the electroweak sphaleron
\index{sphaleron} effects.

\section*{Acknowledgements}
This work was supported by European Union under the RTN
contracts HPRN-CT-2000-00148 and HPRN-CT-2000-00152.

%

\end{document}